\documentclass[review]{amsart}

\usepackage[foot]{amsaddr}

\usepackage{subfig}
\usepackage{graphicx}
\usepackage{caption}
\usepackage{multirow}
\usepackage{longtable}
\usepackage{color,colortbl}
\usepackage{booktabs}
\usepackage{setspace}
\usepackage{amssymb}
\usepackage{amsthm}
 \usepackage{lineno}

\title[Martensite Misorientation Analysis]{Analysis of Misorientation Relationships between Austenite Parents and Twins}

\author{A.F. Brust}
\address[A.F. Brust]{Department of Materials Science and Engineering, The Ohio State University, Columbus, OH 43210, USA}
\author{S.R. Niezgoda}
\address[S.R. Niezgoda]{Department of Materials Science and Engineering, Department of Mechanical and Aerospace Engineering, The Ohio State University, Columbus, OH 43210, USA}
\email[Corresponding author]{niezgoda.6@osu.edu}

\author{V.A. Yardley}
\address[V.A. Yardley]{Eutextikon Computational Materials Consulting LLC, 33 Eastgate St., Stafford ST16 2LZ, U.K. (formerly of Institute for Materials, Ruhr-Universit\"at Bochum, 44780 Bochum, Germany}
\author{E.J. Payton}
\address[E.J. Payton]{Air Force Research Laboratory, Materials and Manufacturing Directorate, Dayton, Ohio 45433, USA}
\thanks{Preprint Submitted to Metallurgical and Materials Transactions A}
\date{April 2, 2018}
\begin{document}

% Abstract
\begin{abstract}
The forward transformation from face centered cubic austenite to body centered cubic/tetragonal martensite in ferrous alloys can significantly influence the microstructure and mechanical properties of the material. Inferring possible high temperature crystal orientations from observations of ambient temperature transformation microstructures is hindered by parent austenite-twin interactions and scatter in the orientation relationship. This creates a major limitation for studying variant selection phenomena and characterizing microstructural response to high temperature thermomechanical processing conditions. In this work, composition tables are developed that detail the product variant boundary misorientation relationships for parent-parent, parent-twin and twin-twin boundary intersections for the Kurdjumov-Sachs (KS), Nishiyama-Wasser\-mann (NW), and an experimentally determined irrational orientation relationship. The frequently referenced KS and NW orientation relationships produce significantly different results from experimental observations. Furthermore, the introduction of a twin into the parent austenite introduces a substantially larger number of misorientation relationships when the orientation relationship is irrational. The effects of crystal symmetry on misorientation results are determined by considering both body centered cubic and body centered tetragonal martensite structures. Lastly, it is observed that some shared variants are found between twins and parents when assuming cubic symmetry but not tetragonal symmetry. The results and relationships may be useful towards accurate and consistent reconstructions of the parent austenite microstructure from observations of martensite.

\end{abstract}

\maketitle
 %\linenumbers


% Introduction
\begin{spacing}{1}
\section{Introduction}

On rapid cooling from the austenite phase field, austenite grains in steels with a carbon content of 0.6\% or below decompose into martensite with 24 crystallographic variants of a lath morphology \cite{Morito,Kitahara_Crystfeat}. At higher concentrations of carbon, a mixture of lath-like and plate-like martensite morphologies are observed \cite{Krauss,Krauss_Marder,Marder_Krauss}. The martensitic transformation is athermal and oftentimes goes to full completion, where little to no parent austenite remains, before the material reaches room temperature. Microstructure characterization must therefore be performed solely on the transformed product, and the microstructure that existed in the austenite phase field must be inferred from observations of the martensitic microstructure. Parent austenite grain structure plays a key role in the performance and properties of the transformed microstructure, such as the ductile to brittle fracture occurrence based on increasing prior austenite grain (PAG) diameter \cite{YardleyEBSD,Kimura} and the classification of creep and cavitation sites \cite{YFP,Hong}.

A transformed austenite grain consists of blocks of laths of two paired crystal variant orientations grouped into packets with a shared habit plane. Both packet and block boundaries are important hindrances to plastic deformation and crack propagation in steels \cite{Matsuda,Guo,MYMH}. Therefore, the size and morphology of the prior austenite grains play a significant role in the mechanical properties of the transformed martensite. Furthermore, the prior austenite structure also contributes to the performance of the material through mechanisms such as impurity segregation at prior austenite grain boundaries, leading to temper embrittlement \cite{Banerji,Horn}. Thus, the reconstruction of the austenite microstructure from the observable martensite is not only highly desired but also necessary for optimizing material processing and performance.

In iron alloys, it is often assumed that the exhibited orientation relationship is close to one of two ``named'' orientation relationships. When approximately 12 variants are observed, the Nishiyama-Wasserman (NW) orientation relationship is referenced \cite{Nishiyama,Wassermann}:

\begin{equation}
    \{111\}_{\gamma}//(011)_{\alpha'};\quad <\bar{1}\bar{1}2>_{\gamma}//<0\bar{1}1>_{\alpha'} \\
\end{equation}
When 24 variants are observed, the Kurdjumov-Sachs (KS) orientation relationship is often cited \cite{KS}:

% KS orientation relationship
\begin{equation}
\centering
    \{111\}_{\gamma}//(011)_{\alpha'}; \quad <\bar{1}01>_{\gamma}//<\bar{1}\bar{1}1>_{\alpha'}
\end{equation}

In the actual orientation relationships, however, the parallel directions and planes are irrational. Cahn and Kalonji \cite{Cahn} argued that, in regards to certain rotations, a symmetry dictated energy extremum exists, although the symmetry does not specify whether the extremum is a minimum, maximum or saddle point. They then determined that no symmetry dictated energy extremum exists for KS, and that the NW orientation relationship produces either a maximum or a shallow minimum. Therefore, both of these cases (KS and NW) can be considered idealized orientation relationships. Deviations from exact parallelism have been characterized by Greninger and Troiano \cite{Greninger} and are predicted by the phenomenological theory of martensitic transformations \cite{WLR,MB1,MB2,MB3,MB4,MB5,Nishiyamabook,Waymanbook}. Experimentally measured orientation relationships will always contain 24 variants but depart from KS or NW orientation relationships. The departure varies with composition and cooling rate \cite{Kitahara_Crystan,Nikravesh}. Anywhere from one to 24 variants can be observed in a prior austenite grain following transformation. Furthermore, it is well known that the martensite crystal structure is body-centered tetragonal (bct) or body-centered cubic (bcc). Which crystal structure is present depends on the concentration of carbon, typically taking the bcc form at less than 0.6\% C, with an expanding c/a ratio with an increasing amount of C \cite{SWLS2006}.

An improved understanding of the variant-to-variant misorientation relationships may be used to improve electron backscatter diffraction (EBSD)-based reconstruction techniques of the prior austenite phase. The prior austenite microstructure can significantly affect physical properties of the phase-transformed material through microstructure scale \cite{Sinha2017} and potentially through crystallographic texture. In the present work, composition tables of the misorientation relationships between variants within a single prior austenite grain and its annealing twins are calculated for three separate orientation relationships: KS, NW, and one which has been experimentally determined \cite{Payton}, building off of the work by Cayron \cite{Cayron}. It is currently not possible to accurately measure the c-axis extension that would distinguish bct from bcc at higher C content using EBSD (resulting in pseudosymmetry in the experimental results). As such, bct martensite in EBSD is typically indexed as bcc. To determine whether this has any impact on the misorientation data, both cubic (point group $m\bar{3}m$) and tetragonal (point group $4/mmm$) symmetry are used to compute the misorientations in standalone composition tables.

% Materials and Methodology section
\section{Materials and Methodology}

% Table 1: Table for the xi and euler angles
\begin{table}[t]
    \centering
    \label{Bunge Euler comparison to xi}
    \begin{tabular}{c c c c}
    \toprule
       Rotation & KS & NW & Exp \\
         \midrule
        $\xi_1$ & $5.26^\circ$ & $0.00^\circ$ & $3.30^\circ$  \\
        $\xi_2$ & $10.30^\circ$ & $9.74^\circ$ & $8.50^\circ$ \\
        $\xi_3$ & $5.26^\circ$ & $9.74^\circ$ & $8.90^\circ$ \\
        
        $\phi_1$ & $114.2^\circ$ & $135.0^\circ$ & $116.1^\circ$ \\
        $\Phi$ & $10.5^\circ$ & $9.6^\circ$ & $8.9^\circ$ \\
        $\phi_3$ & $204.2^\circ$ & $180.0^\circ$ & $200.3^\circ$ \\
        \bottomrule
    \end{tabular}
    \label{tab:my_label}
    \caption{The $\xi$ and Bunge Euler angles for the first variant for the KS, NW and Experimental orientation relationships.}
\end{table}

Starting from a single prior austenite orientation aligned with the sample reference frame (Bunge Euler angles of $\phi_1=\Phi=\phi_2=0)$, the four $\Sigma3$ face-centered cubic annealing twin orientations were calculated from the four unique rotations of $60^\circ$ about $\left<1\,1\,1\right>$.

A number of experimentally observed orientation relationships have been determined from EBSD analysis of SEM and TEM characterized microstructures \cite{Morito,Kitahara_Crystan}. For this paper, the experimentally measured orientation relationship was established based on the procedure outlined in references \cite{YardleyEBSD,YFP,YardleyPayton,Payton} on a sample of low carbon steel, which may be expected to have a KS-like orientation relationship. The orientation relationship was determined through the use of measured $\xi$ angles, with untwinned PAGs being identified and selected manually within a given micrograph range. The datasets were then rotated to coincide the PAG orientation with the sample reference frame, with the angular deviation between the primary axis of the rotation matrix and the axes of the closest Bain correspondence matrix representing the orientation relationship in terms of three parameters. The composition, thermal history, and data collection parameters are published in reference \cite{Natori2005} and the modal orientation relationship values are published in reference \cite{YardleyPayton} and listed in Table 1. Here, $\xi_{1}$ is the smaller deviation from $<110>$-type Bain correspondence axis, $\xi{2}$ is the large deviation from the $<1 1 0>$-type axis and $\xi{3}$ is the deviation from the $<001>$-type axis. This resulted in 24 crystallographic variants, the same number attained with the KS orientation relationship. For convenience of comparison, the three NW and KS modal values and corresponding Bunge Euler angles for a single variant ($V_1$) are also listed in Table 1.

Each of the resulting five orientations (the PAG and its four annealing twins) was then rotated by the 12 $\gamma\to\alpha'$ variant rotations for the NW OR, the 24 $\gamma\to\alpha'$ rotations of the KS OR, or the 24 rotations of an experimentally measured (irrational) orientation relationship. This produced a total of 7260 post-transformation orientations (inclusive of misorientations between identical variants) for the experimental and KS relationships and 1830 orientations for NW. The minimum-angle misorientations between each of these orientations was then calculated assuming either cubic or tetragonal symmetry elements (representing the as-transformed martensite and the tempered martensite ferritic microstructures, respectively). Duplicate misorientation operations were identified to produce the set of potentially observable boundary misorientations within a single twinned PAG, and the unique misorientations were then numbered. Henceforth we will refer to these boundary misorientations as ``intersections.'' We have chosen to represent the variant intersections by a misorientation angle about a specific crystallographic direction. The misorientation that results from a given variant-variant intersection is labeled in a composition table with the cell colored according to the misorientation angle. Additionally, the misorientation axis is plotted in an inverse pole figure. These tables and figures effectively display which misorientation will be observed upon the intersection of two variants.

The numbering of the variants has been conducted in the same manner as in Payton et. al. \cite{Payton} where consecutive variants are grouped into subsets of six that are formed on the same $\{1 1 1\}_{\gamma}$ (i.e., $V_1$ through $V_6$ would share one $\{1 1 1\}_{\gamma}||\{0 1 1\}_{\alpha'}$ relationship, $V_7$ through $V_{12}$ would share a different $\{1 1 1\}_{\gamma}||\{0 1 1\}_{\alpha'}$ relationship and so on).  Additionally, successive variant pairs have the same Bain correspondence matrices ($V_1$ and $V_2$, $V_5$ and $V_6$, $V_9$ and $V_{10}$, etc). Likewise, the numbering of misorientations corresponds to the aforementioned sub-block (low-angle), block and packet boundaries, all differing types of intragranular variant-variant ($V_i-V_j$) interfaces. Misorientations will be denoted as $\Delta g_i$, where $\Delta g_0$ is an identity misorientation operator (identity rotation about an arbitrary or undefined axis to bring the two variants into coincidence with one another) and $\Delta g_1$ corresponds to a sub-block boundary. Block boundaries are distinguished by $\Delta g_{2-4}$, whereas $\Delta g_{5-16}$ refer to packet boundaries. These misorientations all refer to parent-parent cases, where a variant formed from the PAG intersects with another variant that formed from the PAG. For misorientations $\Delta g_{17+}$, which involve twin-variants, the $V_i-V_j$ interfaces are undefined packet boundaries.

% Results Section
\section{Results}

% KS-Cubic Orientation Relationship
\subsection{Kurdjumov-Sachs Orientation Relationship With Cubic Symmetry}

Given a single parent austenite grain transformed to martensite, the composition table assuming the KS orientation relationship is displayed in Figure 1. The two axes in the table correspond to a particular variant, numbered according to Tables 2 through 4, which details the unique misorientation relationships that exist between variants for certain cases (parent-parent, parent-twin and twin-twin). Figure 1 covers the parent-parent case, where variants within a PAG intersect with variants within that same PAG. The plot itself is symmetric, so the top half mirrors the filled in bottom half of the plot. The identity misorientations ($V_i-V_i$) that occur between like variants are unobservable interfaces due to the lack of a misorientation and will later be found outside of the parent-parent case for the KS (and NW) orientation relationships.

% Figure 1: Parent-Parent Aus Composition Table
\begin{figure*}[h]
\centering
%\graphicspath{{./KS/}}
\includegraphics[width=0.6\textwidth]{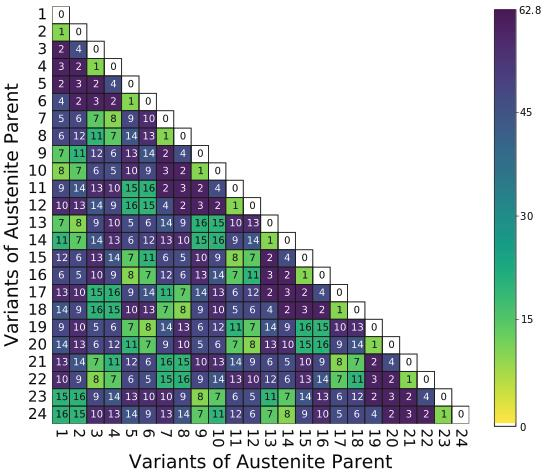}
\caption{Composition table exhibiting the possible variant-variant misorientations for prior parent-parent austenite grains.}
\label{fig1}
\end{figure*}

The boxes in the table corresponding to the variant-variant misorientations are colored by the degree of misorientation, from light to dark, according to the corresponding colorbar. For cubic systems, the maximum misorientation angle is $62.8^{o}$ \cite{Heinz}. The majority of the misorientation squares are dark, meaning the misorientation angle is large for most of the variant intersections. Referring to Table 1, four of the possible 16 misorientations have angles near $60^{o}$, within 10\% of the maximum possible cubic value. Table 2 reveals that there exist only 16 misorientations that occur from variant intersections out of 276 possibilities for the KS orientation relationship (excluding the $V_i-V_i$ cases). However, it is also known that the austenite grains may contain FCC annealing twins before the phase transformation to martensite occurs \cite{Christian}, which would result in twin-related variants. Thus, it is necessary to determine what unique misorientations--if any--would result with the intersection of a parent austenite variant and a twinned austenite variant.

For sake of space and redundancy, only one parent-twin interaction will be presented since the results for all four produce identical sets of misorientations (although the misorientations appear in different locations within the composition table). The full composition tables are available as supplemental data to the present paper. The parent-$60^\circ [1 1 1]$ twin composition table can be found in Figure 2, with corresponding misorientation keys in Table 3.

% Figure 2: Parent-Twin Composition Table
\begin{figure*}[h]
\centering
%\graphicspath{{./KS/}}
\includegraphics[width=0.55\textwidth]{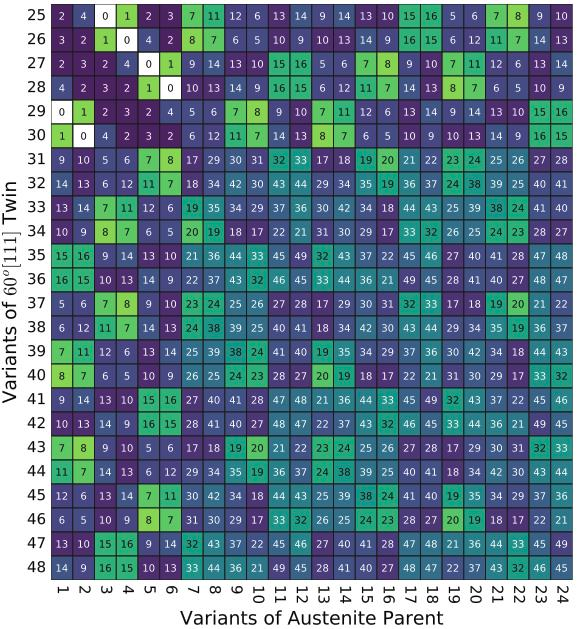}
\caption{The possible variant-variant misorientations for prior parent-$60^\circ [1 1 1]$ twin austenite grains.}
\label{fig2}
\end{figure*}

As can be seen from Figure 2, the parent-twin interactions introduce considerably more misorientations than the parent-parent case alone, with 49 in total being observed. Additionally, the 16 misorientations found within the parent-parent case are repeated in the parent-twin cases, leaving 33 unique misorientations that would indicate the presence of at least one twin within the parent austenite grain. Again, we can see that the intersection squares all tend to be darker, indicative of higher misorientation angles, especially with regards to the unique parent-twin misorientations. The existence of six identity misorientations, consistent for each respective twinning case, results in the possibility of no observable interface within the transformed microstructure where an annealing twin boundary once existed for the KS orientation relationship. 

Although rarer than $\Sigma3$ annealing twins, it is possible for twin-twin variant intersections ($\Sigma9$ boundaries \cite{Randal}) to occur in the parent austenite microstructure. This would consist of the intersection of two variants that transformed from austenite twins of differing rotations that nucleated within the same parent austenite grain; i.e. the former boundary between two different twins of the same parent austenite. Composition tables were constructed for these interactions, with the composition table for the intersections between $60^\circ\left[\bar{1} \bar{1} 1\right]$ and $60^\circ\left[1 1 1\right]$ twins from the same austenite grain shown in Figure 3. The corresponding list of misorientation angle-axis pairs can be found in Table 4. The same misorientations result from all other twin-twin composition tables; the complete set of tables is available in the supplemental material.

% Figure 3: Twin-Twin Composition Table
\begin{figure*}[h]
\centering
%\%graphicspath{{./KS/}}
\includegraphics[width=0.55\textwidth]{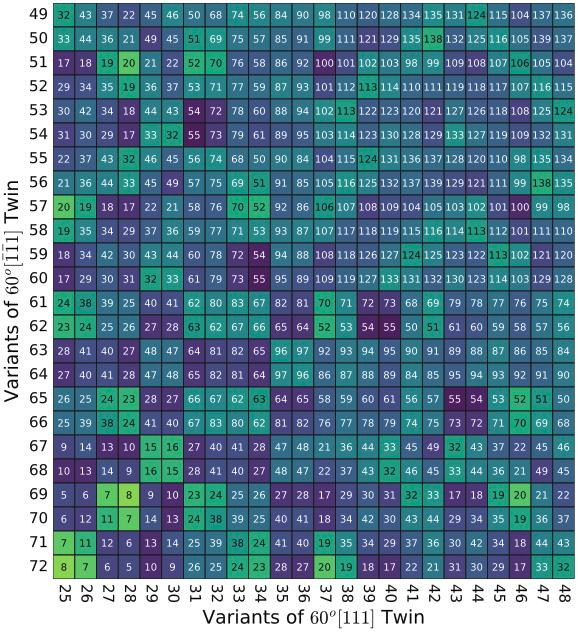}
\caption{Composition table exhibiting the possible variant-variant misorientations for differing twins that nucleated from the same parent austenite grain.}
\label{fig3}
\end{figure*}

The twin-twin case includes misorientations $\Delta g_5$ to $\Delta g_{139}$, indicating a large number of the misorientations found with the parent-parent and parent-twin cases will also appear in twin-twin variant intersections. Additionally, we see that no identity misorientations exist in the twin-twin cases. The twin-twin intersections bring about 90 unique misorientations that can only be observed in twin-twin intersections, and are thus indicative of a prior $\Sigma9$ boundary. 

The existence of misorientations numbered above 16 would necessitate that at least one of the variants being observed came from an austenite twin. Twin variants of the same rotation intersecting with each other (for example, the composition table of $60^\circ\left[\bar{1} \bar{1} 1\right]$ and $60^\circ\left[\bar{1} \bar{1} 1\right]$) produce the exact same composition table as Figure 1. Since these cases are highly unlikely, it can be assumed that the observance of $\Delta g_1$ through $\Delta g_4$ would indicate that at least one of the variants would have had to have nucleated from a parent austenite grain. These tables can also be found in the supplemental material. Additionally, out of all possible misorientations, only three could be considered as low-angle: $\Delta g_1 (10.53^\circ)$, $\Delta g_7 (14.88^\circ)$, and $\Delta g_8 (10.53^\circ)$. The minimum misorientation angles are $\Delta g_1$ and $\Delta g_8$, both $(10.53^\circ)$, and the maximum misorientation angle is $\Delta g_{55} (60.83^\circ)$.

To better visualize the directional aspects of the misorientations with respect to cubic symmetry, the axes for the parent-parent, parent-twin and twin-twin cases were plotted on stereographic triangles and displayed in Figure 4. Tables 2 to 4 can be used to identify the misorientation angles corresponding to each respective misorientation axis. Misorientation axes are colored according to misorientation angle using the same color key as used in Figure 1. Several misorientations are found to exhibit the same axes as one another. We can also see from the tables that some of the misorientation angles are very similar. This will be addressed in the Discussion section of this paper.

% Figure 4: Stereographic Triangles
\begin{figure*}[ht]
\centering
%\graphicspath{{./KS/}}
\subfloat[Misorientation axes for intra-parent case]{\includegraphics[width=0.385\textwidth]{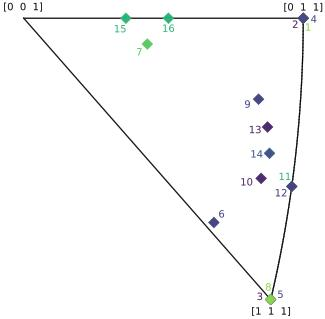}\label{KS_PP_ST}}
\subfloat[Misorientation axes for parent-twin case]{\includegraphics[width=0.4\textwidth]{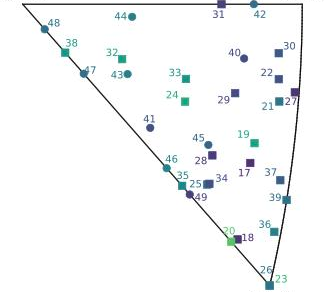}\label{KS_PT_ST}}
\\
\subfloat[Half of the misorientation axes for the twin-twin case]{\includegraphics[width=0.4\textwidth]{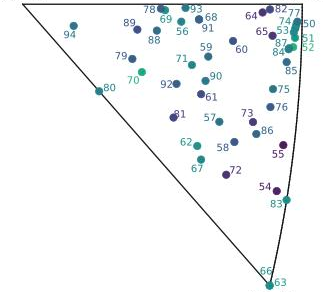}\label{KS_TT1_ST}}
\subfloat[Remaining misorientation axes for the twin-twin case]{\includegraphics[width=0.4\textwidth]{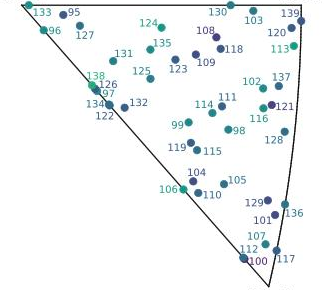}\label{KS_TT2_ST}}
\caption{Misorientations directional axes overlayed on stereographic triangles for the KS orientation relationship for: \protect\subref{KS_PP_ST} the intra-parent case, \protect\subref{KS_PT_ST} the parent-twin case, and \protect\subref{KS_TT1_ST} and \protect\subref{KS_TT2_ST} the twin-twin case (split into two subfigures to reduce the density of points).}
\label{fig4}
\end{figure*}

\subsection{Nishiyama-Wassermann Orientation Relationship}

NW has half of the number of variants as compared to KS due to the alignment of symmetry operators between parent and product phases. As such, $V_1$ and $V_2$ in the KS orientation relationship refer to $V_1$ in the NW orientation relationship, $V_3$ and $V_4$ in KS are $V_2$ in NW, and so on. Due to the reduced number of orientation relationship variants, far fewer unique misorientations can be observed and a complete composition table exhibiting all of the possible variant combinations can be simultaneously represented in Figure 5.

%Figure 5: Full NW Composition Table
\begin{figure*}[h]
\centering{}
%\graphicspath{{./NW/}}
\includegraphics[width=0.9\textwidth]{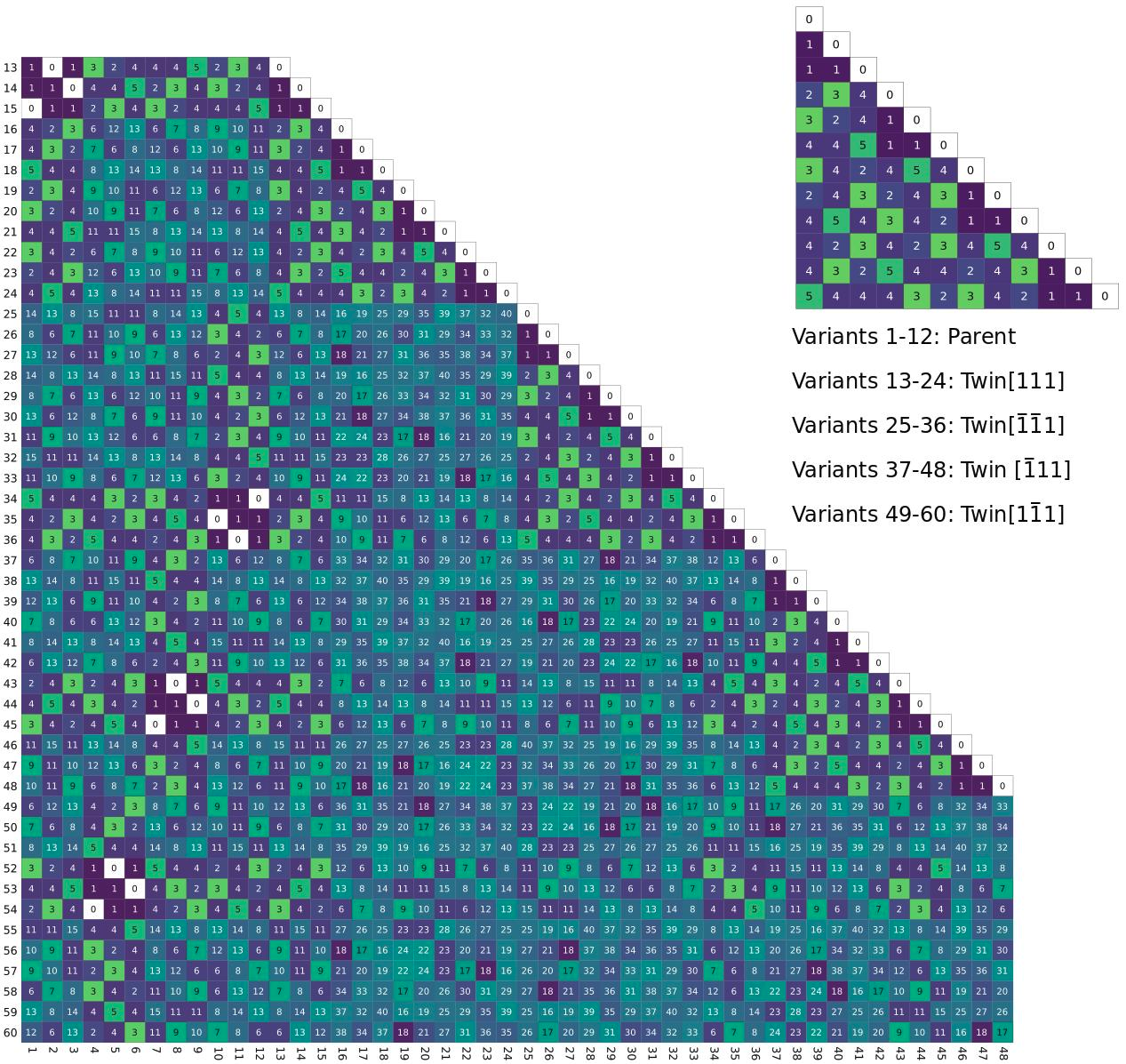}
\caption{Full composition table derived from NW orientation relationship. The subfigure at the upper right can be identified within the full table and comprises the truncated portions at the upper left and lower right portions of the table, representing misorientations derived from parent-parent variant intersections. The key relates the variant numbers to the parent or twin type of the grain the martensite transformed from.}
\label{fig:Full CT NW}
\end{figure*}

The parent-parent case applying the NW orientation relationship involves five unique misorientations, while the parent-twin case has 15 misorientations and includes all of the parent-parent misorientations. Additionally, we can see the existence of three identity operators instead of six as in the KS case. Finally, in terms of the differing twin-twin variant intersections, there exist $\Delta g_2$ through $\Delta g_{40}$, excluding only $\Delta g_1$. In comparing the tables for KS and NW, it is apparent that NW $\Delta g_1$ seems to combine KS $\Delta g_2$-$\Delta g_3$, in the process eliminating the existence of KS $\Delta g_1$. As expected, the NW orientation relationship also results in almost all high misorientation angles, with one possible low-angle boundary, $\Delta g_3 (13.76^\circ)$  as opposed to three in the KS case. Table 5 in the Appendix section lists all of the respective NW misorientations with the corresponding angle-axis pairing. There were no misorientations with axes or angles within $1^\circ$ of those of another misorientation within the NW orientation relationship. The misorientation axes for each case are plotted on the stereographic triangle in Figure 6.
%
% Figure 6: Full NW Stereographic Triangle
\begin{figure*}[h]
\centering{}
%\graphicspath{{./NW/}}
\includegraphics[width=0.35\textwidth]{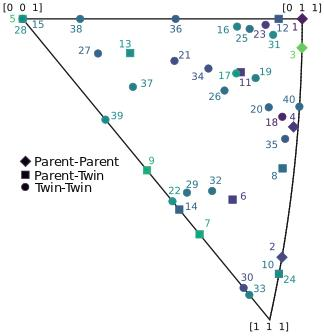}
\caption{Stereographic triangle plotting all misorientation axes with colored points corresponding to degree of misorientation angle for the NW orientation relationship.}
\label{fig:Full CT NW}
\end{figure*}

\subsection{Experimentally Observed Orientation Relationship}

The composition table for the parent-parent case of the experimentally observed (i.e., irrational) orientation relationship is given in Figure 7. Interestingly, calculation resulted in exactly the same misorientation numbering (with only 16 unique misorientations) as Figure 1, but with small deviations in the misorientation angles and/or directional axes from KS. The same color scale is used in Figure 6 as in Figure 1, such that direct comparison of the colors in each box illustrates the misorientation angle differences. The misorientation numbering and the respective angle-axis pairing for each misorientation is given in Tables 6-8. 

% Figure 6: Exp parent-parent composition table
\begin{figure*}[h]
\centering{}
%\graphicspath{{./Experimental/}}
\includegraphics[width=0.5\textwidth]{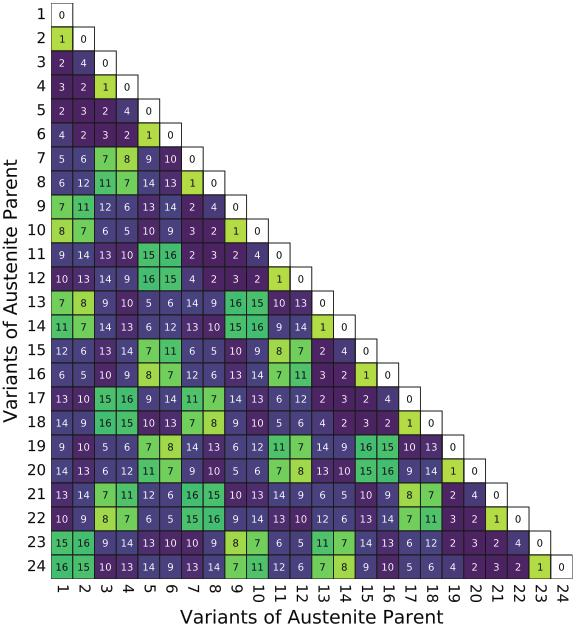}
\caption{Composition table exhibiting parent-parent variant intersections given an experimentally determined orientation relationship.}
\label{fig:Exp_Par-Twin}
\end{figure*}

The parent-twin case was analyzed next, with resultant composition table displayed in Figure 8. Two noticeable differences between the KS and experimental cases are readily observed when comparing Figure 7 to Figure 2. First, the identity operators found in the KS orientation relationship (for example, between $V_30$ and $V_2$ in Figure 2) are replaced by a low angle misorientation, $\Delta g_{19}$ ($3.19 ^\circ [3 3 7]$). The second main difference is the number of misorientations present, increasing from 49 total in the KS case to 71 misorientations using an experimentally measured orientation relationship, resulting in 22 extra misorientations. This difference can be partially explained by the fact that only $\Delta g_2$ from the parent-parent case is observed in the parent-twin case, with all other misorientations being unique to the parent-twin case. In comparison, all sixteen misorientation operators in KS shared between the parent and the twin.

% Figure 7: Exp parent-parent composition table
\begin{figure*}[h]
\centering{}
%\graphicspath{{./Experimental/}}
\includegraphics[width=0.55\textwidth]{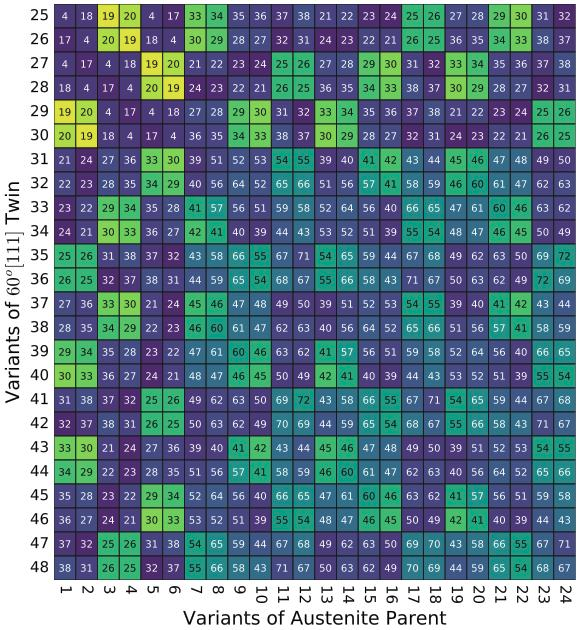}
\caption{Composition table displaying parent-twin variant intersections given an experimentally determined orientation relationship.}
\label{fig:Exp Par-Twin}
\end{figure*}

The twin-twin composition table for the $60^\circ[111]$-$60^\circ[\bar{1}\bar{1}1]$ twins using the experimental orientation relationship is given in Figure 9. The remaining composition tables can be found in the supplemental material. Similar to the parent-twin case, several misorientations exist in addition to the comparative KS case and the only misorientation to exist in both the twin-twin case and the parent-twin or parent-parent cases is $\Delta g_4$. All others are unique to the twin-twin case. There exist 156 unique misorientations for the twin-twin case, with a total of 227 misorientations for the experimental orientation relationship as a whole. This is significantly higher than the 139 misorientations found for the KS orientation relationship. Eight low-angle boundaries exist within the experimental orientation relationship. Three are at the same variant-variant intersection numbers as the KS case, meaning that they are produced from the same variant pairings: $\Delta g_1 (6.60^\circ)$, $\Delta g_7 (12.58^\circ)$, and $\Delta g_8 (8.12^\circ)$. There are five additional low angle misorientations: $\Delta g_{19} (3.19^\circ)$, $\Delta g_{20} (6.93^\circ)$, $\Delta g_{30} (11.15^\circ)$ and $\Delta g_{33} (13.14^\circ)$ unique to the parent-twin case, and $\Delta g_{95} (14.07^\circ)$, unique to the twin-twin case.

% Figure 8: Exp parent-parent composition table
\begin{figure*}[h]
\centering{}
%\graphicspath{{./Experimental/}}
\includegraphics[width=0.55\textwidth]{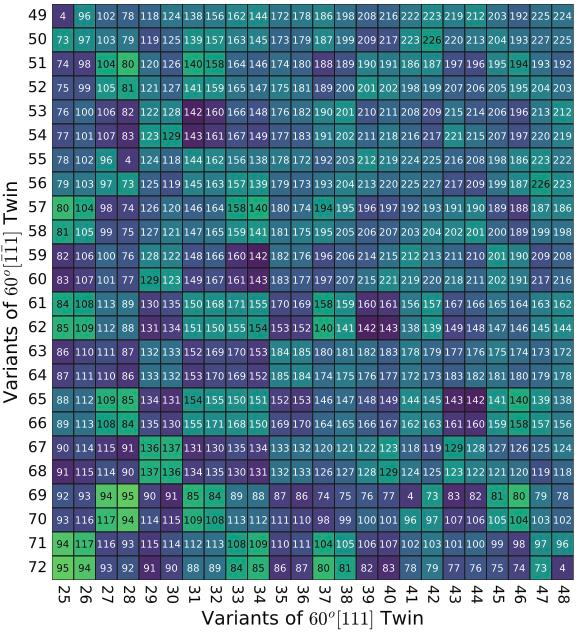}
\caption{Composition table showing differing twin-twin variant intersections given an experimentally determined orientation relationship.}
\label{fig:Exp_Twin-Twin}
\end{figure*}

Stereographic triangles of the misorientation axes corresponding to Figures 6-8 are  given in Figure 10. It is clear from the figure that the experimentally observed orientation relationship tends to produce several misorientations with rotation axes that are essentially parallel, where several of the points are very close to one another in the plot. The colors again relate to the degree of the misorientation by the same color scale as given in Figure 1, while the marker shapes indicate whether the $\Delta g$ originates from the parent-parent, parent-twin, or twin-twin composition tables. 

% Figure 9: Experimental OR Stereographic Triangles
\begin{figure*}[h]
\centering
%\graphicspath{{./Experimental/}}
\subfloat[Misorientation directional axes for the parent-parent case]{\includegraphics[width=0.35\textwidth]{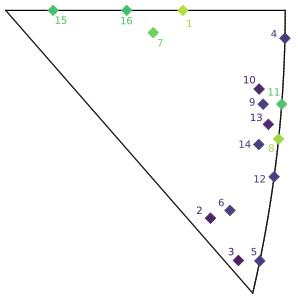}\label{KS_PP_ST}}
\subfloat[Misorientation directional axes for the parent-twin case]{\includegraphics[width=0.35\textwidth]{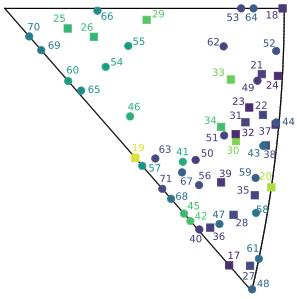}\label{KS_PT_ST}}
\\
\subfloat[One-half of the misorientation axes for the twin-twin case ]{\includegraphics[width=0.35\textwidth]{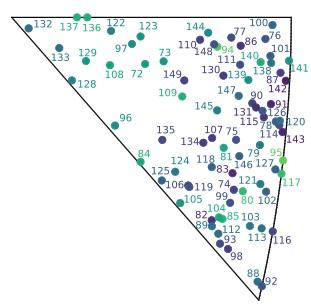}\label{KS_TT1_ST}}
\subfloat[Remainder of misorientation axes for the twin-twin case]{\includegraphics[width=0.35\textwidth]{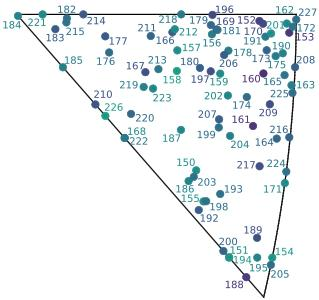}\label{KS_TT2_ST}}
\caption{Misorientations directional axes overlayed on stereographic triangles for the experimentally-measured orientation relationship for: \protect\subref{KS_PP_ST} the parent-parent case, \protect\subref{KS_PT_ST} the parent-twin case, and \protect\subref{KS_TT1_ST} and \protect\subref{KS_TT2_ST} the twin-twin case (split into two subfigures to reduce density of points).}
\label{fig:Stereographic_triangles_KS}
\end{figure*}

\section{KS Orientation Relationship Considering Tetragonal Based Crystal Symmetry}
As mentioned above, the previous cases analyzing $V_i-V_j$ intersections from an orientation relationship standpoint were all conducted using cubic symmetry. That is, the $austenite\to martensite$ transformation was really an $fcc\to bcc$ phase transformation. To study whether product crystal structure affects misorientation data, tetragonal symmetry was applied to the variant rotation matrices in the calculation of the composition tables. This corresponds to the $fcc\to bct$ transformation. The KS orientation relationship was used to compare the effects of cubic and tetragonal symmetry on misorientation calculations. Furthermore, parent-parent, parent-twin and twin-twin variant intersections were examined. The parent-parent composition table is given in Figure 11. The colormapping of the composition tables was comparable in style to the cubic case but was normalized to the maximum tetragonal misorientation angle of $98.42^\circ$ \cite{Heinz} rather than $62.3^\circ$ for cubic systems, as indicated adjacent to the plot. 

% Figure 10: Parent-Parent Tetragonal Composition Table
\begin{figure*}[h]
\centering{}
%\graphicspath{{./Tet/}}
\includegraphics[width=0.65\textwidth]{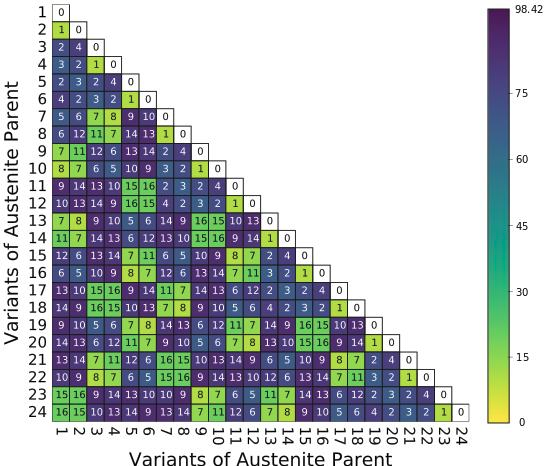}
\caption{Parent-parent composition table for tetragonal symmetry using KS orientation relationship.}
\label{fig:PAus-PAus_KS}
\end{figure*}

In Figure 10, we again see exactly the same misorientation locations within the composition table as both the cubic KS and experimental case, with 16 misorientations in total. Overall, the tetragonal-based misorientation angles seem to fall further away from the maximum misorientation angle. Additionally, as seen in the cubic case, the same three numbered misorientations corresponding to the parent-parent case could be classified as low-angle boundaries: $\Delta g_1 (10.53^\circ)$, $\Delta g_7 (14.88^\circ)$, and $\Delta g_8 (10.53^\circ)$. It is interesting to note that all three of these misorientation numbers were low-angle boundaries across orientation relationship and cubic symmetry when considering 24 variants. All the rest of the misorientations correspond to high-angle boundaries.

% Figure 11: Parent-Twin Tetragonal Composition Table
\begin{figure*}[h]
\centering{}
%\graphicspath{{./Tet/}}
\includegraphics[width=0.55\textwidth]{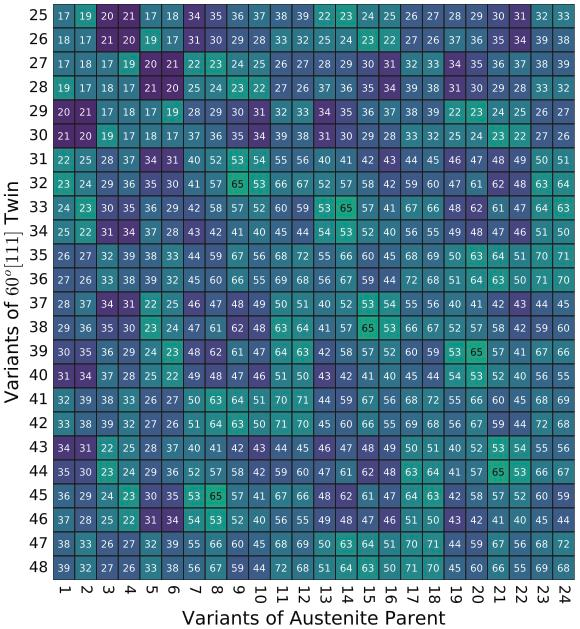}
\caption{Composition table exhibiting the possible variant-variant intersections for prior parent-twin austenite grains using tetragonal symmetry.}
\label{fig:PAus-PTwin_tet}
\end{figure*}

 When considering the parent-twin case for tetragonal symmetry for all four possible twin rotations, all of the misorientation angles are rather large, with none falling below $40^\circ$. Again, we see that there are no shared variants, but in this case, the location where the shared variants exist in the cubic KS case (section 3) is not remotely close to being a low angle, coming in at $90.00^\circ$ (M3). Not only is this misorientation angle very far from an identity rotation, it could not be classified as low angle like the comparable experimental cubic case. No misorientations from the parent-parent case overlap with the parent-twin case, indicating that each misorientation is unique. A total of 55 unique misorientations exist, adding 16 relative to the cubic-KS case. The corresponding composition table is displayed in Figure 12.
 
 % Figure 12: Differing Twin-Twin Tetragonal Composition Table
\begin{figure*}[h]
\centering{}
%\%graphicspath{{./Tet/}}
\includegraphics[width=0.55\textwidth]{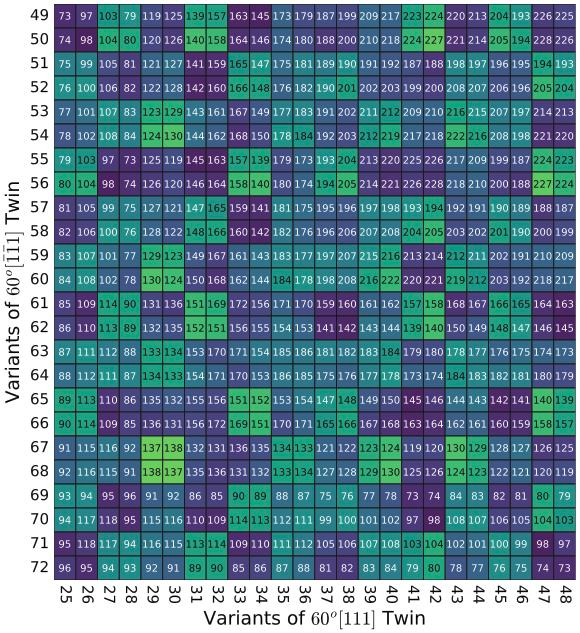}
\caption{Composition table exhibiting the possible variant-variant intersections for differing twin-twin PAGs using tetragonal symmetry.}
\label{fig:PAus-PAeus_KS}
\end{figure*}
 
 The total number of misorientations for the entire tetragonal system increases to 228, substantially more than the 139 found specific to the KS cubic symmetric case. There also seems to be a much larger range of misorientation angle distributions than the parent-twin case, and again no misorientations overlap from either the parent-parent or parent-twin case. It can be concluded that, theoretically, the introduction of tetragonal symmetry to the KS orientation relationship produces purely unique misorientations with regards to the three observed cases of parent-parent, parent-twin or twin-twin variant intersections. 
 
 \section{Discussion}
 
 For the results given, an angular tolerance in radians was taken to four decimal places for the entire $V_i-V_i$ intersection list. The unique misorientations were then taken from this list and given in the composition tables and corresponding tables within the appendix. However, varying the number of angular decimal places does result in some differences in the misorientation angles, usually within a hundredth of a degree, and can also affect the number of unique intersections observed. For example, if too few decimal places were used, more misorientations are deemed to be identical and the total number of unique intersections is reduced. Four places were chosen because this produced produced consistent, realistic results that agreed with prior work by Payton et al\cite{Payton}.
 
 Experimental error in orientation measurement with EBSD will result in some misorientations being indistinguishable from one another. Bingham \textit{et al.} \cite{Bingham} found that 99\% of intragranular orientation measurements within a well annealed grain structure fell within $0.91^\circ$ of one another. Taking $1^\circ$ as a conservative estimate of the angular resolution tolerance for EBSD, it is found in our results that the cubic KS orientation relationship would give three indistinguishable misorientations, as given in Table 11. Two exist in the twin-twin case ($\Delta g_{50}$-$\Delta g_{77}$, $\Delta g_{84}$-$\Delta g_{87}$) and one ($\Delta g_{39}$-$\Delta g_{136}$) exists in both the parent-twin  and the twin-twin tables. The tetragonal KS would give four indistinguishable misorientation operators, where again one misorientation could be misinterpreted within the parent-twin case and the twin-twin case ($\Delta g_{62}$-$\Delta g_{225}$) and three could be misidentified within twin-twin table ($\Delta g_{139}$-$\Delta g_{166}$, $\Delta g_{142}$-$\Delta g_{163}$, $\Delta g_{173}$-$\Delta g_{176}$). Finally, our investigated experimental orientation relationship would give three possible indistinguishable misorientations, where two overlap between the parent-twin and the twin-twin tables ($\Delta g_{59}$-$\Delta g_{102}$, $\Delta g_{61}$-$\Delta g_{205}$) and one within only the twin-twin case ($\Delta g_{147}$-$\Delta g_{174}$). Cubic NW would give zero indistinguishable misorientation operators.
 
 This observation could be significant for a number of reasons. When considering Table 11, some of the possible indistinguishable misorientations stem from variant intersections of $\Sigma3$ twins in the parent grain while others stem from variant intersections of former $\Sigma9$ twins, with none being specific to the parent-parent case. Experimentally, it may be difficult to distinguish between $\Sigma3$ and $\Sigma9$ boundaries if some of the misorientation angle-axis pairings are so similar to each other. In terms of austenite reconstruction codes, this could possibly result in misclassifications of certain variant-variant intersections and thus suggest a parent-twin boundary segment where a twin-twin boundary segment should exist (or vice-versa). It is also worth noting that the scatter in crystallographic orientations for any given variant within the prior austenite grain is typically significantly larger than the experimental error in EBSD \cite{YardleyPayton}.
 
 The repetition of misorientation operators between the parent-parent and parent-twin composition tables may have a significant impact on possible austenite reconstructions, because it means the position of the boundary itself is ambiguous in the cubic-KS orientation relationship. Furthermore, for the parent-twin case, the cubic-KS orientation relationship results in an (unobservable) identity misorientation. This would manifest itself in the transformed microstructure as the former twin boundary appearing discontinuous. The KS-like experimental orientation relationship delivers a misorientation here, but at $\sim3.2^\circ$ it is smaller than a typical threshold for boundary identification in EBSD.
 
 Although not shown in the present work, analysis of several other experimental orientation relationships resulted in similar misorientation distributions with varying misorientation angle-axis pairings. This paper compared the KS and experimental KS-like orientation relationships to show that specific orientation relationships must be applied to differing samples of steel if an accurate analysis of the material is to be constructed. For example, the KS misorientation $\Delta G_1$ has an angle axis pairing of $10.53^\circ@<110>$ whereas the experimental orientation relationship misorientation $\Delta G_1$ exhibits an angle axis pairing of $6.59^\circ@<047>$. Thus, it is clear that the sub-block boundary is substantially different between KS and the experimentally observed orientation relationship, even though the latter is similar to the rational KS orientation relationship. If a separate orientation relationship is used that differs vastly from the KS orientation relationship, it can be assumed that the observable misorientations will differ even more. This would suggest a substantial impact on the accuracy and efficacy of reconstructed austenite microstructures when disparate steel samples are being analyzed. The present work illustrates the potential importance of measuring the orientation relationship in each alloy for reconstruction, since the actual (irrational) misorientations between variants can be significantly different from those in the KS or NW orientation relationships. 
 
 Not only does the orientation relationship itself have a considerable impact on the possible misorientations that can exist between variants generated from an austenite grain, but symmetry plays a major role as well. It is known that we cannot measure the actual orientation when assuming a tetragonal structure due to pseudosymmetry, and as such the common practice is to assume cubic symmetry and neglect any tetragonal c-axis distortion. However, the present work demonstrates that differences exist between the sets of misorientation operators one would expect for the KS orientation relationship when the crystal symmetry is tetragonal as opposed to cubic. First, the latter produces significantly more misorientations than the cubic case (228 compared with 139, respectively). Furthermore, the tetragonal case does not produce any identity misorientations between the parent-twin case, as does the cubic case. In fact, the parent-twin table for tetragonal symmetry does not produce a misorientation angle $< 50.0^\circ$ whereas the cubic case contains 8 unique cases where the misorientation angle is $<30^\circ$, including the case where no misorientation angle exists. The tetragonal composition table would not only suggest that variants adjacent to austenite twin boundaries should not overlap with parent boundaries, but that they should be easily distinguishable for austenite reconstruction codes if tetragonal symmetry could be applied during indexing. Since practical limitations of camera and Hough transform resolution result in better indexing using cubic symmetry, it may be the case that austenite reconstructions may exhibit larger errors with increasing carbon content (tetragonality).

\section{Conclusions}

Both orientation relationship and martensite crystal structure significantly affect possible martensite variant intersections, introducing varying numbers of misorientations and degrees of misorientation. Furthermore, inclusion of prior austenite twins increases the total number of possible misorientations between intersecting variants. From the present work, the following conclusions were drawn:

\begin{enumerate}

\item If the KS or NW orientation relationships were exactly exhibited in a material, then the prior location of an austenite annealing twin boundary would be ambiguous on observation of the product martensite phase due to the presence of identity operators and intra-parent misorientation operators.
 
\item Although experimentally observed orientation relationships in Fe alloys are irrational, the number of misorientations exhibited within a single prior austenite orientation are the same as the KS case (16).

\item The presence of certain characteristic misorientations can be indicative of the presence of a $\Sigma 3$ or $\Sigma 9$ boundary in a prior austenite grain; however, the large number of these possible characteristic misorientations and their similarity to other misorientations that could be exhibited within a single prior austenite grain present a challenge in uniquely identifying the location of the boundaries related to prior austenite annealing twins.

 \end{enumerate}
 
 The results presented here may be useful in austenite reconstruction, as they provide constraints on how an austenite grain could have transformed given the observable martensite.

 \section{Acknowledgements}
 AFB and SRN received support from the Air Force Office of Scientific Research (AFOSR) Summer Faculty Fellowship Program (SFFP) for the portion of this work performed at the Materials and Manufacturing Directorate of the Air Force Research Laboratory (AFRL/RX) and from the Dayton Area Graduate Studies Institute (DAGSI) for the portion of the work performed at Ohio State University. EJP was supported by the Deutsche Forschungsgemeinschaft (DFG) for the portion of this work completed at the Federal Institute for Materials Research and Testing (BAM) in Berlin (grant PA 2285/1-1) and by the Metallic Materials and Processes Research Team for the portion performed at AFRL/RX. VAY was supported under DFG grant YA 326/2-1 for the portion of the work performed at Ruhr-Universit\"at Bochum.
%Start the appendix section which will contain tables related to misorientations that don't necessarily need to be included above

%\onecolumn
\pagebreak
% Section containing misorientation information
\section{Appendices}
\appendix
For Tables 2-11 below, which list the complete misorientation list for the analyzed cases, a few of the notations may be new to the reader and as such will be described here briefly. Consistent with the text, $\Delta g_i$ refers to a specific misorientation resulting from the variant-variant interactions. The term $\theta (\Delta g_i)$ refers to the misorientation angle, always measured in degrees, while the term $\overrightarrow{r} \equiv [r_1,r_2,r_3]$ is the approximate low-index axis of rotation for the misorientation. Finally, the term $\delta (\Delta g_i)$ is the deviation of the true axis from the true axis from the approximate low-index axis of rotation. Finally, the concluding table--Table 12--is the comparison of similar misorientations that fall within $1^\circ$ of each other. In regards to notation, the differing misorientations are denoted by the subscripts (i,j) such that $\Delta \theta (\Delta g_{i,j})$ would represent the angular difference between misorientation angles $\theta (\Delta g_i)$ and $\theta (\Delta g_j)$. Additionally, $\Delta \overrightarrow{r} (\Delta g_{i,j})$ would represent the angular difference between misorientation axes $r_i$ and $r_j$.

\section{KS-Cubic Orientation Relationship Misorientation Data}
%
% Table A.1 for misorientation list parent-parent KS-cubic
\begin{longtable}{ l  c  c  c }
\caption{Parent-parent misorientation list for KS orientation relationship considering cubic symmetry.}\\
%\small
\\
\toprule
\textbf{$\Delta g_i$}&\textbf{$\theta (\Delta g_i)$ ($^\circ$)}&\textbf{$\overrightarrow{r}$}&\textbf{$\delta (\Delta g_i)$ ($^\circ$)} \\
\midrule
	0 &  0.00 & [0 0 1] & 0.00 \\ 
	1 & 10.53 & [0 1 1] & 0.25 \\ 
	2 & 60.00 & [0 1 1] & 0.25 \\ 
	3 & 60.00 & [1 1 1] & 0.00 \\ 
	4 & 49.27 & [0 1 1] & 0.25 \\ 
	5 & 49.27 & [1 1 1] & 0.00 \\ 
	6 & 50.42 & [2 2 3] & 1.39 \\ 
	7 & 14.90 & [1 3 8] & 2.93 \\ 
	8 & 10.53 & [1 1 1] & 0.00 \\ 
	9 & 50.51 & [1 4 5] & 1.89 \\ 
	10 & 57.21 & [3 5 6] & 0.28 \\ 
	11 & 20.60 & [3 5 5] & 1.79 \\ 
	12 & 51.73 & [3 5 5] & 1.79 \\ 
	13 & 57.21 & [2 5 6] & 0.73 \\ 
	14 & 47.11 & [2 4 5] & 2.28 \\ 
	15 & 20.60 & [0 1 3] & 1.27 \\ 
	16 & 21.06 & [0 3 7] & 0.96 \\ 
	\bottomrule
\end{longtable}
 
% Table A.2 for misorientation list for parent-twin KS-cubic
\begin{longtable}{ c c  c  c}
\caption{Parent-twin misorientation list for KS orientation relationship considering cubic symmetry.}\\
\\
\toprule
\textbf{$\Delta g_i$} & \textbf{$\theta (\Delta g_i)$ ($^\circ$)} & \textbf{$\overrightarrow{r}$} & \textbf{$\delta (\Delta g_i)$ ($^\circ$)} \\ 
\midrule
\endfirsthead
\multicolumn{4}{@{}l}{\ldots continued}\\\midrule
\textbf{$\Delta g_i$} & \textbf{$\theta (\Delta g_i)$ ($^\circ$)} & \textbf{$\overrightarrow{r}$} & \textbf{$\delta (\Delta g_i)$ ($^\circ$)} \\ 
\midrule
\endhead % all the lines above this will be repeated on every page
\midrule
\multicolumn{4}{r@{}}{continued \ldots}\\
\endfoot
\endlastfoot
%$textbf{$\Delta g_i$} & \textbf{$\theta (\Delta g_i)$ ($^\circ$)} & \textbf{$\overrightarrow{r}$} & \textbf{$\delta (\Delta g_i)$ ($^\circ$)} \\ 
%\midrule
	17 & 55.61 & [3 5 6] & 1.05 \\ 
	18 & 54.84 & [3 3 4] & 1.33 \\ 
	19 & 25.00 & [2 4 5] & 0.99 \\ 
	20 & 15.45 & [3 3 4] & 0.63 \\ 
	21 & 37.24 & [2 6 7] & 1.60 \\ 
	22 & 47.56 & [2 6 7] & 2.78 \\ 
	23 & 21.06 & [1 1 1] & 0.00 \\ 
	24 & 23.51 & [1 2 4] & 1.45 \\ 
	25 & 40.28 & [3 4 6] & 2.35 \\ 
	26 & 38.94 & [1 1 1] & 0.00 \\ 
	27 & 55.23 & [1 4 4] & 1.34 \\ 
	28 & 53.51 & [2 3 5] & 2.26 \\ 
	29 & 49.19 & [2 5 7] & 1.27 \\ 
	30 & 45.80 & [1 6 7] & 1.20 \\ 
	31 & 51.32 & [0 5 8] & 0.74 \\ 
	32 & 26.11 & [1 2 7] & 0.70 \\ 
	33 & 29.12 & [1 3 6] & 2.27 \\ 
	34 & 47.56 & [3 4 6] & 2.10 \\ 
	35 & 33.57 & [1 1 2] & 1.71 \\ 
	36 & 34.85 & [4 5 5] & 1.15 \\ 
	37 & 44.35 & [3 5 5] & 1.45 \\ 
	38 & 28.41 & [1 1 8] & 0.00 \\ 
	39 & 38.94 & [2 3 3] & 0.76 \\ 
	40 & 49.19 & [1 5 7] & 1.22 \\ 
	41 & 47.83 & [3 3 8] & 2.17 \\ 
	42 & 40.28 & [0 4 5] & 0.92 \\ 
	43 & 34.85 & [1 2 6] & 1.33 \\ 
	44 & 35.45 & [0 1 3] & 1.81 \\ 
	45 & 43.00 & [2 3 5] & 0.48 \\ 
	46 & 33.57 & [1 1 2] & 1.71 \\ 
	47 & 40.28 & [1 1 5] & 1.35 \\ 
	48 & 38.94 & [1 1 9] & 3.65 \\ 
	49 & 51.80 & [4 4 7] & 0.00 \\ 
	\bottomrule
	\end{longtable}

% Table A.3 for misorientation list for twin-twin KS-cubic
\begin{longtable}{ c  c  c  c  |c  c  c  c}
\caption{Twin-twin misorientation list for KS orientation relationship considering cubic symmetry.}\\
\\
\toprule
\textbf{$\Delta g_i$} & \textbf{$\theta (\Delta g_i)$ ($^\circ$)} & \textbf{$\overrightarrow{r}$} & \textbf{$\delta (\Delta g_i)$ ($^\circ$)}  & \textbf{$\Delta g_i$} & \textbf{$\theta (\Delta g_i)$ ($^\circ$)} & \textbf{$\overrightarrow{r}$} & \textbf{$\delta (\Delta g_i)$ ($^\circ$)} \\ 
\midrule
\endfirsthead
\multicolumn{8}{@{}l}{\ldots continued}\\\midrule
\textbf{$\Delta g_i$} & \textbf{$\theta (\Delta g_i)$ ($^\circ$)} & \textbf{$\overrightarrow{r}$} & \textbf{$\delta (\Delta g_i)$ ($^\circ$)}  & \textbf{$\Delta g_i$} & \textbf{$\theta (\Delta g_i)$ ($^\circ$)} & \textbf{$\overrightarrow{r}$} & \textbf{$\delta (\Delta g_i)$ ($^\circ$)} \\ 
\midrule
\endhead

\multicolumn{8}{r@{}}{continued \ldots}\\
\endfoot
\endlastfoot

	50 & 38.33 & [0 1 1] & 3.21  & 95 & 45.10 & [0 1 8] & 1.44 \\ 
	51 & 27.82 & [1 6 6] & 2.59  & 96 & 31.59 & [1 1 9] & 3.65 \\ 
	52 & 21.84 & [1 6 6] & 1.75  & 97 & 33.25 & [2 2 9] & 0.52 \\ 
	53 & 32.33 & [1 6 6] & 3.28  & 98 & 33.57 & [1 2 3] & 1.65 \\ 
	54 & 59.55 & [3 5 5] & 1.84  & 99 & 33.25 & [2 3 6] & 1.79 \\ 
	55 & 60.83 & [1 2 2] & 2.04  & 100 & 57.94 & [5 5 6] & 0.39 \\ 
	56 & 33.75 & [0 1 2] & 2.40  & 101 & 49.43 & [3 4 4] & 1.73 \\ 
	57 & 38.63 & [1 2 3] & 0.67  & 102 & 32.33 & [2 6 7] & 1.46 \\ 
	58 & 45.54 & [3 5 7] & 0.73  & 103 & 36.93 & [0 4 5] & 1.02 \\ 
	59 & 39.87 & [1 4 7] & 1.25  & 104 & 47.91 & [4 4 7] & 2.19 \\ 
	60 & 46.75 & [1 5 7] & 1.73  & 105 & 38.94 & [4 5 7] & 1.00 \\ 
	61 & 48.70 & [2 4 7] & 1.49  & 106 & 28.05 & [4 4 7] & 1.45 \\ 
	62 & 30.26 & [2 3 5] & 0.69  & 107 & 35.43 & [4 5 5] & 1.85 \\ 
	63 & 28.41 & [1 1 1] & 0.00  & 108 & 51.73 & [1 5 8] & 1.73 \\ 
	64 & 55.06 & [0 4 5] & 1.17  & 109 & 48.08 & [1 4 7] & 0.77 \\ 
	65 & 54.92 & [1 6 7] & 1.98  & 110 & 42.11 & [4 4 7] & 1.30 \\ 
	66 & 31.59 & [1 1 1] & 0.00  & 111 & 43.13 & [1 2 3] & 1.65 \\ 
	67 & 33.25 & [3 4 7] & 1.27  & 112 & 40.28 & [5 5 6] & 0.00 \\ 
	68 & 39.87 & [0 4 7] & 2.11  & 113 & 26.45 & [1 6 6] & 1.75 \\ 
	69 & 29.87 & [0 3 7] & 0.92  & 114 & 33.57 & [1 2 3] & 1.81 \\ 
    70 & 24.01 & [1 2 5] & 1.71 &  115 & 38.94 & [3 4 7] & 0.41 \\ 
    71 & 33.75 & [1 3 6] & 1.71 &  116 & 30.93 & [2 5 6] & 0.79 \\ 
    72 & 55.06 & [3 4 6] & 1.21 &  117 & 44.08 & [4 5 5] & 1.40 \\ 
    73 & 53.85 & [2 4 5] & 1.66 &  118 & 45.54 & [1 5 8] & 1.06 \\ 
    74 & 31.96 & [0 1 1] & 2.82 &  119 & 43.40 & [3 4 7] & 1.54 \\ 
    75 & 37.94 & [2 6 7] & 1.23 &  120 & 44.26 & [0 1 1] & 3.21 \\  
    76 & 45.10 & [2 6 7] & 1.37 &  121 & 50.75 & [2 6 7] & 0.83 \\ 
    77 & 38.63 & [0 1 1] & 2.37 &  122 & 44.73 & [2 2 7] & 1.12 \\ 
    78 & 42.85 & [0 2 5] & 0.64 &  123 & 43.87 & [1 3 6] & 1.22 \\ 
    79 & 44.35 & [1 2 6] & 1.17 &  124 & 28.05 & [0 3 7] & 3.23 \\ 
    80 & 34.36 & [2 2 9] & 0.52 &  125 & 35.43 & [1 2 5] & 0.00 \\ 
    81 & 50.75 & [2 3 6] & 1.19 &  126 & 45.80 & [2 2 9] & 0.00 \\ 
    82 & 49.43 & [0 6 7] & 0.90 &  127 & 38.94 & [0 1 6] & 3.04 \\ 
    83 & 31.59 & [2 3 3] & 0.76 &  128 & 40.28 & [2 5 5] & 1.54 \\ 
    84 & 35.78 & [1 6 7] & 2.57 &  129 & 47.91 & [2 3 3] & 2.51 \\ 
    85 & 41.21 & [1 5 5] & 2.15 &  130 & 36.92 & [0 2 3] & 0.91 \\ 
    86 & 42.11 & [2 4 5] & 0.00 &  131 & 35.43 & [1 2 6] & 0.65 \\ 
    87 & 35.11 & [1 6 7] & 2.61 &  132 & 44.08 & [2 2 7] & 1.75 \\ 
    88 & 39.99 & [1 3 8] & 3.46 &  133 & 30.93 & [0 0 1] & 1.15 \\ 
    89 & 46.95 & [1 3 8] & 3.26 &  134 & 40.28 & [1 1 4] & 1.39 \\ 
    90 & 36.93 & [1 3 5] & 1.04 &  135 & 31.96 & [1 3 8] & 1.01 \\ 
    91 & 42.85 & [0 4 7] & 2.26 &  136 & 38.33 & [2 3 3] & 0.15 \\ 
    92 & 44.35 & [1 2 4] & 1.55 &  137 & 41.21 & [2 6 7] & 2.31 \\ 
    93 & 36.93 & [0 1 2] & 0.89 &  138 & 22.75 & [2 2 9] & 0.83 \\ 
    94 & 37.94 & [0 1 7] & 3.18 &  139 & 45.80 & [0 1 1] & 1.94 \\ 
    \bottomrule
\end{longtable}

\section{NW-Cubic Orientation Relationship Misorientation Data}
% Table B.4 for NW orientation relationship with cubic symmetry
\begin{longtable}{ c  c  c  c  |c  c c  c }
 \caption{Misorientation list for NW orientation relationship with cubic symmetry.}\\
\toprule
\textbf{$\Delta g_i$} & \textbf{$\theta (\Delta g_i)$ ($^\circ$)} & \textbf{$\overrightarrow{r}$} & \textbf{$\delta (\Delta g_i)$ ($^\circ$)}  & \textbf{$\Delta g_i$} & \textbf{$\theta (\Delta g_i)$ ($^\circ$)} & \textbf{$\overrightarrow{r}$} & \textbf{$\delta (\Delta g_i)$ ($^\circ$)} \\ 
\midrule
\endfirsthead
\multicolumn{8}{@{}l}{\ldots continued}\\\midrule
\textbf{$\Delta g_i$} & \textbf{$\theta (\Delta g_i)$ ($^\circ$)} & \textbf{$\overrightarrow{r}$} & \textbf{$\delta (\Delta g_i)$ ($^\circ$)}  & \textbf{$\Delta g_i$} & \textbf{$\theta (\Delta g_i)$ ($^\circ$)} & \textbf{$\overrightarrow{r}$} & \textbf{$\delta (\Delta g_i)$ ($^\circ$)} \\ 
\midrule
\endhead

\multicolumn{8}{r@{}}{continued \ldots}\\
\endfoot
\endlastfoot
	0 &  0.00 & [0 0 1] & 0.00 & 21 & 48.11 & [1 4 8] & 1.45 \\ 
	1 & 60.00 & [0 1 1] & 0.25 & 22 & 30.01 & [1 1 2] & 0.49 \\ 
	2 & 50.05 & [3 4 4] & 0.36 & 23 & 52.24 & [0 5 6] & 0.85 \\ 
	3 & 13.76 & [1 6 6] & 3.32 & 24 & 31.59 & [5 6 6] & 0.18 \\ 
	4 & 53.69 & [1 3 3] & 0.78 & 25 & 38.61 & [0 3 4] & 1.36 \\ 
	5 & 19.47 & [0 0 1] & 0.00 & 26 & 39.12 & [1 4 6] & 1.31 \\ 
	6 & 51.41 & [2 3 4] & 1.19 & 27 & 42.40 & [1 2 9] & 1.44 \\ 
	7 & 24.47 & [3 3 5] & 0.00 & 28 & 31.59 & [0 0 1] & 0.00 \\ 
	8 & 40.66 & [1 2 2] & 2.34 & 29 & 37.58 & [1 1 2] & 2.37 \\ 
	9 & 23.12 & [2 2 5] & 0.27 & 30 & 49.12 & [5 5 6] & 0.00 \\ 
	10 & 38.94 & [5 6 6] & 0.18 & 31 & 31.70 & [0 6 7] & 1.97 \\ 
	11 & 52.63 & [1 5 7] & 0.65 & 32 & 38.94 & [3 4 6] & 1.18 \\ 
	12 & 45.38 & [0 6 7] & 0.86 & 33 & 35.43 & [5 5 6] & 1.07 \\ 
	13 & 30.75 & [1 3 9] & 1.44 & 34 & 46.72 & [1 5 8] & 1.24 \\ 
	14 & 42.69 & [1 1 2] & 1.15 & 35 & 45.38 & [1 3 3] & 2.05 \\ 
	15 & 38.94 & [0 0 1] & 0.00 & 36 & 43.37 & [0 1 2] & 1.04 \\ 
	16 & 33.56 & [0 5 7] & 0.88 & 37 & 35.43 & [1 2 6] & 0.58 \\ 
	17 & 27.47 & [1 5 7] & 0.00 & 38 & 38.94 & [0 1 6] & 0.26 \\ 
	18 & 58.94 & [1 3 3] & 2.74 & 39 & 31.48 & [1 1 4] & 0.61 \\ 
	19 & 34.92 & [1 5 6] & 1.53 & 40 & 41.08 & [1 4 4] & 0.43 \\ 
	20 & 41.76 & [2 6 7] & 1.34 &  {} & {} & {} & {}  \\ 
	\bottomrule
\end{longtable}

\section{Experimental-Cubic Orientation Relationship Misorientation Data}
% Table C.5 for Experimental OR with Cubic Symmetry considering parent-parent variants
\begin{longtable}{ c  c  c  c }
\caption{Parent-parent misorientation list for experimental orientation relationship considering cubic symmetry.}\\
\toprule
\textbf{$\Delta g_i$} & \textbf{$\theta (\Delta g_i)$ ($^\circ$)} & \textbf{$\overrightarrow{r}$} & \textbf{$\delta (\Delta g_i)$ ($^\circ$)} \\ 
 \midrule
 \endfirsthead
\multicolumn{4}{@{}l}{\ldots continued}\\\midrule  
\textbf{$\Delta g_i$} & \textbf{$\theta (\Delta g_i)$ ($^\circ$)} & \textbf{$\overrightarrow{r}$} & \textbf{$\delta (\Delta g_i)$ ($^\circ$)} \\ 
\midrule
\endhead
\multicolumn{4}{r@{}}{continued \ldots}\\
\endfoot
\endlastfoot
    0 &  0.00 & [0 0 1] & 0.00 \\ 
    1 &  6.60 & [0 4 7] & 0.36 \\ 
    2 & 59.48 & [2 2 3] & 2.56 \\ 
    3 & 60.14 & [5 5 6] & 1.73 \\ 
    4 & 53.70 & [1 6 6] & 3.16 \\ 
    5 & 52.51 & [4 5 5] & 1.94 \\ 
    6 & 51.85 & [3 4 5] & 0.95 \\ 
    7 & 12.58 & [0 3 7] & 3.47 \\ 
    8 &  8.12 & [2 5 5] & 0.77 \\ 
    9 & 52.31 & [2 6 7] & 1.64 \\ 
    10 & 58.63 & [2 6 7] & 2.30 \\ 
    11 & 16.32 & [1 3 3] & 1.29 \\ 
    12 & 51.54 & [1 2 2] & 1.32 \\ 
    13 & 57.59 & [2 5 5] & 2.11 \\ 
    14 & 51.55 & [2 4 5] & 3.33 \\ 
    15 & 17.00 & [0 1 7] & 0.00 \\ 
    16 & 17.80 & [0 3 8] & 0.00 \\  
    \bottomrule
\end{longtable}

% Table C.6 for Experimental OR with cubic symmetry considering parent-twin variants
\begin{longtable}{ c  c c  c | c  c  c  c }
\caption{Parent-twin misorientation list for experimental orientation relationship applying cubic symmetry.}\\
\toprule
\textbf{$\Delta g_i$} & \textbf{$\theta (\Delta g_i)$ ($^\circ$)} & \textbf{$\overrightarrow{r}$} & \textbf{$\delta (\Delta g_i)$ ($^\circ$)}  & \textbf{$\Delta g_i$} & \textbf{$\theta (\Delta g_i)$ ($^\circ$)} & \textbf{$\overrightarrow{r}$} & \textbf{$\delta (\Delta g_i)$ ($^\circ$)} \\ 
\midrule
\endfirsthead
\multicolumn{8}{@{}l}{\ldots continued}\\\midrule
\textbf{$\Delta g_i$} & \textbf{$\theta (\Delta g_i)$ ($^\circ$)} & \textbf{$\overrightarrow{r}$} & \textbf{$\delta (\Delta g_i)$ ($^\circ$)}  & \textbf{$\Delta g_i$} & \textbf{$\theta (\Delta g_i)$ ($^\circ$)} & \textbf{$\overrightarrow{r}$} & \textbf{$\delta (\Delta g_i)$ ($^\circ$)} \\ 
\midrule
\endhead

\multicolumn{8}{r@{}}{continued \ldots}\\
\endfoot
\endlastfoot

    17 & 57.28 & [5 5 6] & 0.85 & 45 & 18.68 & [3 3 5] & 0.66  \\ 
    18 & 53.62 & [0 1 1] & 0.25  & 46 & 22.31 & [2 3 7] & 1.85  \\ 
    19 &  3.19 & [3 3 7] & 0.87  & 47 & 40.86 & [4 5 6] & 2.17  \\ 
    20 &  6.93 & [3 5 5] & 0.80  & 48 & 41.83 & [1 1 1] & 0.44  \\ 
    21 & 51.90 & [1 5 6] & 2.47 & 49 & 53.72 & [1 5 6] & 2.57  \\ 
    22 & 50.74 & [1 3 3] & 2.39  & 50 & 53.47 & [3 4 6] & 2.29  \\ 
    23 & 57.03 & [2 6 7] & 0.83  & 51 & 50.56 & [2 4 5] & 1.71  \\ 
    24 & 58.13 & [1 5 5] & 0.82  & 52 & 47.49 & [1 6 6] & 1.71  \\ 
    25 & 19.93 & [0 1 5] & 3.01  & 53 & 49.65 & [0 4 5] & 0.58  \\ 
    26 & 20.11 & [1 2 8] & 2.94  & 54 & 26.00 & [1 2 7] & 1.36  \\ 
    27 & 49.57 & [1 1 1] & 2.99  & 55 & 26.91 & [1 3 8] & 1.42  \\ 
    28 & 48.79 & [4 5 6] & 0.22  & 56 & 48.56 & [3 4 6] & 1.10  \\ 
    29 & 15.54 & [0 3 7] & 1.73  & 57 & 29.33 & [3 3 7] & 1.08  \\ 
    30 & 11.15 & [2 4 5] & 0.32  & 58 & 37.83 & [2 3 3] & 1.68  \\ 
    31 & 51.66 & [2 5 6] & 0.67  & 59 & 43.70 & [3 5 5] & 2.72  \\  
    32 & 57.66 & [2 4 5] & 0.63  & 60 & 26.83 & [1 1 5] & 0.67  \\  
    33 & 13.14 & [1 4 5] & 1.54  & 61 & 38.86 & [4 5 5] & 1.89  \\  
    34 & 17.90 & [1 2 3] & 2.50  & 62 & 50.90 & [1 5 7] & 1.50  \\ 
    35 & 50.99 & [3 5 5] & 2.33  & 63 & 49.68 & [3 4 8] & 2.43  \\  
    36 & 51.70 & [3 3 4] & 2.06  & 64 & 44.99 & [0 6 7] & 0.18  \\  
    37 & 54.61 & [2 5 5] & 1.26  & 65 & 32.05 & [2 2 9] & 0.44  \\ 
    38 & 48.54 & [2 5 5] & 2.29  & 66 & 32.23 & [0 2 7] & 0.44  \\  
    39 & 54.35 & [2 3 4] & 1.24  & 67 & 43.91 & [3 4 7] & 2.23  \\  
    40 & 52.78 & [2 2 3] & 0.90  & 68 & 37.88 & [4 4 7] & 0.27  \\  
    41 & 23.38 & [3 4 7] & 0.88  & 69 & 36.70 & [1 1 9] & 0.00  \\ 
    42 & 17.38 & [2 2 3] & 0.80  & 70 & 36.30 & [1 1 9] & 3.10  \\ 
    43 & 40.42 & [2 5 5] & 2.58  & 71 & 49.81 & [1 1 2] & 1.15  \\ 
    44 & 46.50 & [1 3 3] & 1.40  & {} & {} & {} & {} \\
    \bottomrule 
\end{longtable}

% Table C.7 for Experimental OR using cubic symmetry for twin-twin variants
\begin{longtable}{ c  c  c  c|  c  c  c  c }
\caption{Twin-twin misorientation list for experimental orientation relationship using cubic symmetry.}\\
\toprule
\textbf{$\Delta g_i$} & \textbf{$\theta (\Delta g_i)$ ($^\circ$)} & \textbf{$\overrightarrow{r}$} & \textbf{$\delta (\Delta g_i)$ ($^\circ$)}  & \textbf{$\Delta g_i$} & \textbf{$\theta (\Delta g_i)$ ($^\circ$)} & \textbf{$\overrightarrow{r}$} & \textbf{$\delta (\Delta g_i)$ ($^\circ$)} \\ 
\midrule
\endfirsthead
\multicolumn{8}{@{}l}{\ldots continued}\\\midrule
\textbf{$\Delta g_i$} & \textbf{$\theta (\Delta g_i)$ ($^\circ$)} & \textbf{$\overrightarrow{r}$} & \textbf{$\delta (\Delta g_i)$ ($^\circ$)}  & \textbf{$\Delta g_i$} & \textbf{$\theta (\Delta g_i)$ ($^\circ$)} & \textbf{$\overrightarrow{r}$} & \textbf{$\delta (\Delta g_i)$ ($^\circ$)} \\ 
\midrule
\endhead

\multicolumn{8}{r@{}}{continued \ldots}\\
\endfoot
\endlastfoot

 72 & 27.85 & [1 3 7] & 1.03 & 150 & 31.29 & [3 4 7] & 1.39 \\   
 73 & 29.15 & [1 4 8] & 0.97 & 151 & 31.47 & [4 4 5] & 0.00 \\ 
 74 & 51.26 & [2 3 4] & 1.42 & 152 & 52.44	& [0 5 6] & 0.89 \\ 
 75 & 47.47 & [2 4 5] & 1.50 & 153 & 53.97 & [0 1 1] & 2.56 \\ 
 76 & 46.40 & [0 6 7] & 2.86 & 154 & 28.76 & [4 5 5] & 0.82 \\ 
 77 & 48.93 & [0 5 7] & 2.77 & 155 & 30.95 & [3 3 5] & 2.44 \\ 
 78 & 43.57 & [1 3 3] & 1.60 & 156 & 36.40	& [0 5 8] & 2.74 \\ 
 79 & 37.51 & [2 5 6] & 2.51 & 157 & 30.75 & [1 4 8] & 1.21 \\ 
 80 & 19.19 & [3 5 6] & 1.68 & 158 & 26.18 & [1 3 6] & 0.84 \\ 
 81 & 25.33 & [3 5 7] & 1.83 & 159 & 31.83 & [1 4 6] & 0.53 \\ 
 82 & 54.79 & [4 5 7] & 2.03 & 160 & 55.66 & [1 5 6] & 0.71 \\ 
 83 & 56.60 & [2 3 4] & 1.38 & 161 & 54.33 & [2 5 6] & 0.36 \\ 
 84 & 24.58 & [2 2 5] & 0.64 & 162 & 33.20 & [0 1 1] & 1.82 \\ 
 85 & 21.43 & [3 4 5] & 2.23 & 163 & 35.86 & [1 5 5] & 0.82 \\ 
 86 & 54.57 & [1 5 7] & 3.46 & 164 & 42.00 & [2 5 5] & 1.93 \\ 
 87 & 52.32 & [1 6 6] & 1.30 & 165 & 38.51 & [1 5 5] & 1.64 \\ 
 88 & 38.94 & [1 1 1] & 2.18 & 166 & 44.43 & [0 1 2] & 2.58 \\ 
 89 & 38.07 & [2 2 3] & 1.92 & 167 & 47.93 & [1 3 7] & 1.72 \\ 
 90 & 51.18 & [2 6 7] & 1.37 & 168 & 31.80 & [1 1 3] &	0.50 \\ 
 91 & 57.28 & [2 6 7] & 2.09 & 169 & 49.11 & [0 2 3] & 2.07 \\ 
 92 & 46.63 & [1 1 1] & 1.57 & 170 & 49.94 & [0 5 6] & 1.21 \\ 
 93 & 48.62 & [3 3 4] & 1.23 & 171 & 31.95 & [3 5 5] & 1.67 \\ 
 94 & 15.89 & [1 5 7] & 3.10 & 172 & 37.96 & [0 1 1] & 1.54 \\ 
 95 & 14.07 & [1 2 2] & 1.24 & 173 & 42.28 & [1 6 7] & 1.51 \\ 
 96 & 31.95 & [2 2 7] & 1.70 & 174 & 40.90 & [1 3 4] & 1.03 \\ 
 97 & 32.02 & [1 3 8] & 2.86 & 175 & 36.08 & [1 6 7] & 1.73 \\ 
 98 & 52.02 & [3 3 4] & 1.52 & 176 & 41.13 & [1 2 8] & 2.09 \\ 
 99 & 47.61 & [3 4 5] & 1.00 & 177 & 45.83 & [0 1 4] & 3.34 \\ 
100 & 42.19 & [0 6 7] & 1.90 & 178 & 36.78 & [1 5 7] & 1.77 \\ 
101 & 44.68 & [1 6 7] & 1.85 & 179 & 41.48 & [0 5 8] & 1.34 \\ 
102 & 43.07 & [3 5 5] & 1.86 & 180 & 43.80	& [1 4 7] & 0.96 \\ 
103 & 37.37 & [4 5 6] & 1.33 & 181 & 38.66 & [0 2 3] &	2.20 \\ 
104 & 24.19 & [3 4 5] & 2.56 & 182 & 40.20	& [0 1 7] & 0.58 \\ 
105 & 29.99 & [4 4 7] & 1.05 & 183 & 44.60	& [0 1 9] & 2.23 \\ 
106 & 50.59 & [3 4 6] & 2.88 & 184 & 34.28 & [0 0 1] &	0.00 \\ 
107 & 52.39 & [1 2 3] & 0.90 & 185 & 34.90	& [1 1 7] & 0.46 \\ 
108 & 26.20 & [1 2 7] & 1.11 & 186 & 33.91	& [3 4 7] & 2.48 \\ 
109 & 22.16 & [1 2 4] & 2.62 & 187 & 33.16 & [2 3 6] &	1.50 \\ 
110 & 51.14 & [0 3 5] & 3.09 & 188 & 52.87	& [5 5 6] & 1.81 \\ 
111 & 48.22 & [1 5 7] & 1.94 & 189 & 47.82	& [4 5 6] & 2.46 \\ 
112 & 40.84 & [3 3 4] & 2.48 & 190 & 34.51	& [1 6 7] & 2.77 \\ 
113 & 38.94 & [3 4 4] & 2.01 & 191 & 36.76	& [0 6 7] & 2.93 \\ 
114 & 47.76 & [1 3 3] & 1.16 & 192 & 43.93 & [3 3 5] & 1.20 \\ 
115 & 54.01 & [2 6 7] & 1.08 & 193 & 38.94 & [4 5 7] & 1.03 \\ 
116 & 50.41 & [3 4 4] & 1.17 & 194 & 28.83 & [4 4 5] & 0.45 \\ 
117 & 19.13 & [1 2 2] & 0.00 & 195 & 33.86 & [4 5 5] & 2.47 \\ 
118 & 42.58 & [3 5 7] & 1.68 & 196 & 49.33 & [0 5 8] & 0.24 \\ 
119 & 48.29 & [3 4 6] & 2.71 & 197 & 47.99 & [1 4 6] & 1.53 \\ 
120 & 39.65 & [1 3 3] & 0.95 & 198 & 39.00 & [3 4 6] & 2.62 \\ 
121 & 36.78 & [3 5 6] & 2.04 & 199 & 38.69 & [1 2 3] & 0.32 \\  
122 & 35.21 & [0 2 7] & 2.10 & 200 & 42.48 & [3 3 4] & 0.41 \\ 
123 & 29.79 & [0 2 5] & 2.65 & 201 & 30.43 & [0 6 7] & 2.58 \\ 
124 & 37.20 & [3 4 7] & 1.69 & 202 & 33.64 & [1 3 4] & 1.77 \\ 
125 & 43.19 & [1 1 2] & 1.45 & 203 & 38.94 & [3 4 7] & 2.02 \\ 
126 & 45.92 & [2 6 7] & 0.74 & 204 & 34.48 & [1 2 3] &	2.20 \\ 
127 & 42.72 & [1 2 2] & 1.05 & 205 & 39.09 & [4 5 5] & 1.71 \\ 
128 & 34.25 & [1 1 6] & 0.74 & 206 & 44.68 & [1 5 7] & 1.73 \\ 
129 & 28.20 & [1 2 9] & 0.00 & 207 & 43.87 & [1 2 3] & 1.23 \\ 
130 & 52.88 & [1 4 6] & 1.31 & 208 & 43.71 & [1 6 6] & 0.05 \\  
131 & 55.81 & [2 6 7] & 0.52 & 209 & 48.36 & [2 6 7] & 1.08 \\ 
132 & 38.94 & [0 0 1] & 3.34& 210 & 46.74 & [1 1 4] & 0.61 \\ 
133 & 39.48 & [1 1 9] & 2.48 & 211 & 42.36 & [0 2 5] & 3.19 \\ 
134 & 52.91 & [1 2 3] & 1.90 & 212 & 31.86 & [0 1 2] & 2.08 \\ 
135 & 49.28 & [2 3 6] & 0.92 & 213 & 37.33 & [1 3 7] & 0.46 \\ 
136 & 22.82 & [0 2 9] & 0.32 & 214 & 45.16 & [0 1 5] & 0.39 \\ 
137 & 22.48 & [0 1 5] & 0.53 & 215 & 38.94 & [0 1 7] & 1.24 \\ 
138 & 37.30 & [1 6 7] & 1.56 & 216 & 42.56 & [2 5 5] & 0.57 \\ 
139 & 31.14 & [1 4 5] & 0.49 & 217 & 47.56 & [3 5 6] & 1.70 \\ 
140 & 23.20 & [1 6 7] & 0.00 & 218 & 37.05 & [0 1 2] & 0.70 \\ 
141 & 29.46 & [1 6 6] & 1.27 & 219 & 35.44 & [1 2 5] &	0.00 \\ 
142 & 59.98 & [1 5 5] & 0.70 & 220 & 41.39 & [2 3 8] & 1.54 \\ 
143 & 58.86 & [1 3 3] & 1.29 & 221 & 32.86 & [0 1 9] & 2.03 \\ 
144 & 32.21 & [0 5 8] & 2.12 & 222 & 36.83 & [1 1 3] & 0.22 \\ 
145 & 34.72 & [2 5 7] & 1.35 & 223 & 32.48 & [1 2 5] & 1.11 \\ 
146 & 43.95 & [2 4 5] & 0.99 & 224 & 38.46 & [1 2 2] & 0.47 \\ 
147 & 40.59 & [1 3 4] & 1.09 & 225 & 39.20 & [1 4 4] & 0.55 \\ 
148 & 47.46 & [1 5 7] & 3.61 & 226 & 27.63 & [2 2 7] & 0.95 \\ 
149	& 50.84 & [1 4 7] & 1.92 & 227 & 40.96 & [0 1 1] & 0.25 \\ 
\bottomrule
\end{longtable}

% Last but not least, let's list the misorientations delivered from the KS OR applying tetragonal symmetry
\section{KS-Tetragonal Orientation Relationship Misorientation Data}

% Table D.8 for parent-parent KS orientation relationship using Tetragonal symmetry
\begin{longtable}{ c  c  c  c }
\caption{Parent-parent misorientation list for KS orientation relationship with respect to tetragonal symmetry.}\\
\toprule
\textbf{$\Delta g_i$} & \textbf{$\theta (\Delta g_i)$ ($^\circ$)} & \textbf{$\overrightarrow{r}$} & \textbf{$\delta (\Delta g_i)$ ($^\circ$)} \\ 
 \midrule
 \endfirsthead
\multicolumn{4}{@{}l}{\ldots continued}\\\midrule  
\textbf{$\Delta g_i$} & \textbf{$\theta (\Delta g_i)$ ($^\circ$)} & \textbf{$\overrightarrow{r}$} & \textbf{$\delta (\Delta g_i)$ ($^\circ$)} \\ 
\midrule
\endhead
\multicolumn{4}{r@{}}{continued \ldots}\\
\endfoot
\endlastfoot
    0 &  0.00 & [1 0 0] & 0.00 \\ 
    1 & 10.53 & [1 0 1] & 0.08 \\ 
    2 & 76.27 & [5 5 1] & 0.82 \\ 
    3 & 70.53 & [1 1 0] & 0.08 \\ 
    4 & 82.82 & [3 3 1] & 0.39 \\ 
    5 & 71.21 & [7 5 0] & 0.57 \\ 
    6 & 77.65 & [7 5 1] & 0.98 \\ 
    7 & 14.88 & [8 3 1] & 2.93 \\ 
    8 & 10.53 & [1 1 1] & 0.76 \\ 
    9 & 90.00 & [6 5 1] & 2.69 \\ 
    10 & 84.26 & [7 6 1] & 3.18 \\ 
    11 & 20.60 & [5 5 3] & 1.79 \\ 
    12 & 83.14 & [5 4 0] & 0.92 \\ 
    13 & 90.00 & [7 6 1] & 2.28 \\ 
    14 & 85.62 & [3 2 0] & 2.28 \\ 
    15 & 20.60 & [3 0 1] & 1.27 \\ 
    16 & 21.06 & [7 3 0] & 0.96 \\ 
    \bottomrule
\end{longtable}

% Table D.9 for parent-twin KS orientation relationship using Tetragonal symmetry
\begin{longtable}{ c  c  c  c|  c  c  c c }
\caption{Parent-twin misorientation list for KS orientation relationship with tetragonal symmetry.}\\
\toprule
\textbf{$\Delta g_i$} & \textbf{$\theta (\Delta g_i)$ ($^\circ$)} & \textbf{$\overrightarrow{r}$} & \textbf{$\delta (\Delta g_i)$ ($^\circ$)}  & \textbf{$\Delta g_i$} & \textbf{$\theta (\Delta g_i)$ ($^\circ$)} & \textbf{$\overrightarrow{r}$} & \textbf{$\delta (\Delta g_i)$ ($^\circ$)} \\ 
\midrule
\endfirsthead
\multicolumn{8}{@{}l}{\ldots continued}\\\midrule
\textbf{$\Delta g_i$} & \textbf{$\theta (\Delta g_i)$ ($^\circ$)} & \textbf{$\overrightarrow{r}$} & \textbf{$\delta (\Delta g_i)$ ($^\circ$)}  & \textbf{$\Delta g_i$} & \textbf{$\theta (\Delta g_i)$ ($^\circ$)} & \textbf{$\overrightarrow{r}$} & \textbf{$\delta (\Delta g_i)$ ($^\circ$)} \\ 
\midrule
\endhead

\multicolumn{8}{r@{}}{continued \ldots}\\
\endfoot
\endlastfoot
    17 & 60.00 & [1 0 1] & 0.08  & 45 & 66.84 & [5 2 3] & 2.59 \\ 
    18 & 60.00 & [1 1 1] & 0.00  & 46 & 79.33 & [4 0 1] & 0.57 \\  
    19 & 49.47 & [1 0 1] & 0.08  & 47 & 70.53 & [6 0 1] & 3.04 \\ 
    20 & 90.00 & [1 0 0] & 0.00  & 48 & 77.17 & [6 1 3] & 1.37 \\  
    21 & 90.48 & [8 0 1] & 0.46  & 49 & 73.21 & [6 0 3] & 0.45 \\ 
    22 & 50.51 & [4 1 5] & 1.89  & 50 & 55.23 & [4 1 4] & 1.34 \\ 
    23 & 47.11 & [5 2 4] & 2.28  & 51 & 56.04 & [7 1 5] & 0.80 \\ 
    24 & 57.21 & [6 2 5] & 0.73  & 52 & 57.28 & [7 2 4] & 1.33 \\ 
    25 & 57.21 & [5 3 6] & 0.28  & 53 & 45.80 & [6 1 7] & 1.20 \\ 
    26 & 70.53 & [9 1 1] & 3.65  & 54 & 51.32 & [5 0 8] & 0.74 \\ 
    27 & 71.21 & [9 1 1] & 1.50  & 55 & 65.68 & [8 1 0] & 2.55 \\ 
    28 & 70.53 & [1 1 1] & 0.00  & 56 & 65.82 & [9 1 2] & 1.22 \\  
    29 & 71.21 & [4 3 3] & 1.28  & 57 & 60.00 & [5 0 3] & 2.46 \\ 
    30 & 76.27 & [9 1 0] & 3.74  & 58 & 65.68 & [3 1 0] & 0.40 \\ 
    31 & 84.26 & [9 0 1] & 0.53  & 59 & 72.16 & [7 0 3] & 2.54 \\ 
    32 & 58.90 & [7 2 4] & 1.44  & 60 & 68.84 & [7 1 4] & 0.72 \\  
    33 & 57.21 & [6 3 5] & 0.28  & 61 & 62.19 & [9 0 1] & 1.52 \\ 
    34 & 85.62 & [8 1 1] & 0.77  & 62 & 80.13 & [6 0 3] & 1.50 \\  
    35 & 77.65 & [9 2 1] & 2.84  & 63 & 49.19 & [7 1 5] & 1.22 \\ 
    36 & 65.82 & [4 1 3] & 1.80  & 64 & 51.73 & [6 0 3] & 2.15 \\ 
    37 & 60.83 & [3 0 2] & 1.32  & 65 & 40.28 & [5 0 4] & 0.92 \\ 
    38 & 67.12 & [5 2 4] & 0.83  & 66 & 58.37 & [9 2 0] & 3.13 \\  
    39 & 68.83 & [8 2 5] & 0.00  & 67 & 57.28 & [7 1 1] & 1.26 \\ 
    40 & 60.00 & [5 1 4] & 0.38  & 68 & 60.00 & [8 1 4] & 1.69 \\ 
    41 & 64.21 & [6 0 5] & 0.68  & 69 & 64.21 & [3 0 1] & 1.00 \\ 
    42 & 73.10 & [9 2 1] & 2.25  & 70 & 51.80 & [5 0 1] & 0.16 \\ 
    43 & 80.13 & [6 1 0] & 0.63  & 71 & 51.32 & [9 1 0] & 2.30 \\ 
    44 & 70.22 & [5 1 2] & 0.87  & 72 & 57.28 & [3 0 2] & 0.91 \\ 
    \bottomrule
\end{longtable}

% Table D.10 for twin-twin KS orientation relationship using Tetragonal symmetry
\begin{longtable}{ c  c  c  c | c  c  c  c }
\caption{Twin-twin misorientation list applying KS orientation relationship with tetragonal symmetry.}\\
\toprule
\textbf{$\Delta g_i$} & \textbf{$\theta (\Delta g_i)$ ($^\circ$)} & \textbf{$\overrightarrow{r}$} & \textbf{$\delta (\Delta g_i)$ ($^\circ$)}  & \textbf{$\Delta g_i$} & \textbf{$\theta (\Delta g_i)$ ($^\circ$)} & \textbf{$\overrightarrow{r}$} & \textbf{$\delta (\Delta g_i)$ ($^\circ$)} \\ 
\midrule
\endfirsthead
\multicolumn{8}{@{}l}{\ldots continued}\\\midrule
\textbf{$\Delta g_i$} & \textbf{$\theta (\Delta g_i)$ ($^\circ$)} & \textbf{$\overrightarrow{r}$} & \textbf{$\delta (\Delta g_i)$ ($^\circ$)}  & \textbf{$\Delta g_i$} & \textbf{$\theta (\Delta g_i)$ ($^\circ$)} & \textbf{$\overrightarrow{r}$} & \textbf{$\delta (\Delta g_i)$ ($^\circ$)} \\ 
\midrule
\endhead

\multicolumn{8}{r@{}}{continued \ldots}\\
\endfoot
\endlastfoot
    73 & 89.37 & [7 1 2] & 0.83  & 151 & 30.26 & [4 3 7] & 0.69 \\ 
    74 & 88.41 & [8 3 1] & 1.01  & 152 & 28.41 & [1 1 1] & 0.00 \\ 
    75 & 55.61 & [6 3 5] & 1.05  & 153 & 67.12 & [5 3 3] & 0.97 \\ 
    76 & 49.19 & [7 2 5] & 1.27  & 154 & 58.16 & [6 4 3] & 1.85 \\ 
    77 & 67.12 & [8 3 4] & 0.44  & 155 & 75.34 & [5 2 0] & 0.75 \\ 
    78 & 73.10 & [7 4 4] & 1.21  & 156 & 78.62 & [7 3 1] & 1.15 \\ 
    79 & 47.56 & [6 2 7] & 2.78  & 157 & 39.87 & [4 0 7] & 2.11 \\ 
    80 & 37.24 & [6 2 7] & 1.60  & 158 & 29.87 & [3 0 7] & 0.92 \\ 
    81 & 82.82 & [6 0 1] & 1.38  & 159 & 88.45 & [7 1 2] & 0.93 \\ 
    82 & 84.80 & [3 0 1] & 2.03  & 160 & 90.00 & [7 1 3] & 0.62 \\  
    83 & 54.84 & [5 4 4] & 1.33  & 161 & 67.84 & [7 5 4] & 1.03 \\ 
    84 & 55.61 & [6 5 3] & 1.05  & 162 & 69.41 & [5 4 2] & 2.38 \\ 
    85 & 81.62 & [8 2 1] & 0.98  & 163 & 92.92 & [5 0 2] & 0.28 \\ 
    86 & 79.33 & [4 1 0] & 0.57  & 164 & 89.07 & [6 3 0] & 1.77 \\ 
    87 & 53.51 & [5 2 3] & 2.22  & 165 & 45.10 & [6 2 7] & 1.20 \\ 
    88 & 63.26 & [7 3 5] & 0.90  & 166 & 38.63 & [1 0 1] & 0.79 \\ 
    89 & 38.94 & [1 1 1] & 0.00  & 167 & 81.13 & [7 3 3] & 0.60 \\ 
    90 & 40.28 & [4 3 6] & 2.35  & 168 & 86.95 & [7 4 3] & 1.11 \\ 
    91 & 68.84 & [5 3 2] & 1.64  & 169 & 34.36 & [2 2 9] & 0.52 \\ 
    92 & 65.82 & [6 5 3] & 1.84  & 170 & 76.43 & [6 3 4] & 2.29 \\ 
    93 & 49.47 & [1 1 1] & 0.00  & 171 & 67.12 & [2 1 1] & 0.49 \\ 
    94 & 50.51 & [3 2 2] & 1.39  & 172 & 80.91 & [5 2 0] & 0.45 \\ 
    95 & 90.00 & [9 1 2] & 3.42  & 173 & 67.69 & [8 2 3] & 2.23 \\ 
    96 & 84.26 & [9 1 0] & 0.53  & 174 & 67.84 & [7 3 2] & 0.77 \\ 
    97 & 84.80 & [8 2 3] & 1.05  & 175 & 63.93 & [6 1 3] & 0.95 \\ 
    98 & 94.25 & [5 1 2] & 0.52  & 176 & 68.01 & [8 2 3] & 2.76 \\ 
    99 & 54.84 & [5 4 4] & 1.33  & 177 & 54.92 & [8 2 1] & 2.61 \\ 
    100 & 47.56 & [6 3 4] & 2.10  & 178 & 46.95 & [8 3 1] & 3.26 \\ 
    101 & 62.19 & [3 1 1] & 1.20  & 179 & 77.00 & [7 2 3] & 1.29 \\ 
    102 & 67.12 & [8 4 3] & 0.44  & 180 & 76.62 & [7 3 3] & 1.28 \\ 
    103 & 44.35 & [5 3 5] & 1.45  & 181 & 54.21 & [7 1 3] & 1.48 \\ 
    104 & 34.85 & [5 4 5] & 1.15  & 182 & 59.08 & [9 2 2] & 0.24 \\ 
    105 & 77.17 & [8 1 2] & 1.49  & 183 & 52.83 & [9 0 1] & 2.76 \\ 
    106 & 80.41 & [7 1 3] & 0.80  & 184 & 45.10 & [8 1 0] & 1.44 \\ 
    107 & 47.56 & [6 4 3] & 2.10  & 185 & 58.60 & [1 0 0] & 2.92 \\ 
    108 & 49.19 & [7 5 2] & 1.27  & 186 & 59.08 & [6 1 0] & 0.96 \\ 
    109 & 90.00 & [4 1 1] & 2.56  & 187 & 75.87 & [5 1 2] & 1.31 \\ 
    110 & 86.85 & [8 2 1] & 2.77  & 188 & 85.47 & [7 1 3] & 0.80 \\ 
    111 & 47.83 & [8 3 3] & 2.17  & 189 & 57.94 & [5 5 6] & 0.39 \\ 
    112 & 57.28 & [6 2 3] & 0.87  & 190 & 49.43 & [4 3 4] & 1.73 \\ 
    113 & 40.28 & [6 4 3] & 2.35  & 191 & 68.27 & [7 1 2] & 1.49 \\ 
    114 & 38.94 & [3 2 3] & 0.76  & 192 & 72.16 & [3 1 1] & 0.22 \\ 
    115 & 63.26 & [5 3 1] & 0.45  & 193 & 47.91 & [4 4 7] & 2.13 \\ 
    116 & 58.90 & [6 5 2] & 1.41  & 194 & 38.94 & [5 4 7] & 1.00 \\ 
    117 & 51.73 & [5 5 3] & 1.79  & 195 & 69.41 & [7 0 2] & 1.13 \\ 
    118 & 84.26 & [4 0 1] & 0.70  & 196 & 71.99 & [7 0 3] & 1.68 \\ 
    119 & 76.27 & [7 4 2] & 2.50  & 197 & 52.60 & [8 3 4] & 0.00 \\ 
    120 & 78.90 & [5 4 1] & 1.64  & 198 & 52.79 & [7 3 2] & 1.19 \\ 
    121 & 67.12 & [5 2 1] & 0.52  & 199 & 77.89 & [7 1 4] & 0.96 \\ 
    122 & 73.10 & [7 3 0] & 3.05  & 200 & 81.76 & [5 4 3] & 1.63 \\ 
    123 & 35.45 & [3 1 0] & 1.81  & 201 & 40.28 & [5 5 6] & 0.00 \\ 
    124 & 29.12 & [6 3 1] & 2.27  & 202 & 73.69 & [5 1 1] & 2.00 \\ 
    125 & 81.62 & [7 3 1] & 0.62  & 203 & 77.14 & [5 2 1] & 0.93 \\ 
    126 & 83.97 & [7 4 1] & 0.72  & 204 & 38.94 & [4 3 7] & 0.41 \\ 
    127 & 62.19 & [6 3 2] & 1.89  & 205 & 30.93 & [5 2 6] & 0.79 \\ 
    128 & 67.12 & [7 4 1] & 0.33  & 206 & 75.34 & [8 0 5] & 0.24 \\ 
    129 & 34.85 & [6 1 2] & 1.33  & 207 & 56.60 & [7 3 2] & 0.08 \\ 
    130 & 26.11 & [7 2 1] & 0.77  & 208 & 58.37 & [8 4 1] & 2.88 \\ 
    131 & 79.33 & [5 4 2] & 0.83  & 209 & 66.00 & [5 2 2] & 2.05 \\ 
    132 & 71.88 & [5 5 2] & 0.93  & 210 & 67.12 & [3 2 1] & 0.99 \\ 
    133 & 38.94 & [9 1 1] & 3.65  & 211 & 50.27 & [3 1 0] & 0.61 \\ 
    134 & 40.28 & [5 1 1] & 1.35  & 212 & 43.87 & [6 3 1] & 1.22 \\ 
    135 & 75.36 & [4 3 1] & 2.00  & 213 & 82.11 & [4 1 1] & 1.21 \\ 
    136 & 83.08 & [3 2 1] & 1.85  & 214 & 82.45 & [8 3 2] & 1.18 \\ 
    137 & 20.60 & [3 1 0] & 1.27  & 215 & 47.91 & [7 0 2] & 0.68 \\ 
    138 & 21.06 & [7 0 3] & 0.96  & 216 & 38.94 & [6 0 1] & 3.04 \\ 
    139 & 38.33 & [1 0 1] & 3.21  & 217 & 67.12 & [6 3 1] & 2.07 \\ 
    140 & 27.82 & [6 1 6] & 2.59  & 218 & 69.63 & [6 4 1] & 0.74 \\ 
    141 & 90.00 & [7 0 2] & 0.53  & 219 & 36.92 & [3 2 0] & 0.91 \\ 
    142 & 92.48 & [7 0 3] & 0.92  & 220 & 86.14 & [8 3 2] & 1.59 \\ 
    143 & 59.55 & [5 5 3] & 1.84 & 221 & 87.31 & [7 4 2] & 0.94 \\ 
    144 & 61.99 & [3 3 1] & 1.40  & 222 & 30.93 & [1 0 0] & 1.15 \\ 
    145 & 93.46 & [7 3 1] & 0.90  & 223 & 40.28 & [1 1 4] & 1.39 \\ 
    146 & 85.91 & [8 4 1] & 1.05  & 224 & 31.96 & [3 1 8] & 1.27 \\ 
    147 & 45.54 & [7 3 5] & 0.73  & 225 & 79.91 & [2 0 1] & 1.04 \\ 
    148 & 39.87 & [7 1 4] & 1.25  & 226 & 83.79 & [5 5 3] & 1.53 \\ 
    149 & 72.16 & [7 4 3] & 2.14  & 227 & 22.75 & [2 2 9] & 0.83 \\ 
    150 & 78.40 & [7 5 3] & 1.77  & 228 & 86.44 & [5 5 2] & 0.65 \\ 
    \bottomrule
\end{longtable}

\section{Similar Misorientation Comparisons}
% KS misorientation angle comparison

% KS  cubic symmetry similar misorientation comparison
\begin{table}[h]
\centering
 \caption{The possible misorientation pairs that may be indistinguishable upon experimental observation. KS-Cubic refers to KS orientation relationship assuming cubic symmetry, Low-C Expt. is the referenced experimental orientation relationship and KS-Tetrag as the KS orientation relationship assuming tetragonal symmetry. }
\label{Redundant Misorientations}
\begin{tabular}{ c  c  c  c c }
\toprule
OR & $\Delta g_i$ & $\Delta g_j$ & $\Delta\theta(\Delta g_{i,j}) (^\circ)$ & $\Delta\overrightarrow{r}(\Delta g_{i,j}) (^\circ)$\\
\midrule
KS-Cubic & 39 & 136 & 0.619 & 0.409 \\
KS-Cubic & 50 & 77 & 0.309 & 0.909 \\
KS-Cubic & 84 & 87 & 0.670 & 0.115 \\
Low-C Expt. & 59 & 102 & 0.630 & 0.772 \\
Low-C Expt. & 61 & 205 & 0.229 & 0.010 \\
Low-C Expt. & 147 & 174 & 0.315 & 0.725 \\
KS-Tetrag. & 62 & 225 & 0.223 & 0.492 \\
KS-Tetrag. & 139 & 166 & 0.309 & 0.909 \\
KS-Tetrag. & 142 & 163 & 0.441 & 0.354 \\
KS-Tetrag. & 173 & 176 & 0.315 & 0.532 \\
\bottomrule
\end{tabular}
\end{table}

\end{spacing}
\end{document}